\documentclass[12pt]{article}

\usepackage{scicite}

\usepackage{graphics}
\usepackage{color}
\usepackage{float}
\usepackage{graphicx}
\usepackage{amssymb}
\usepackage{overpic}
\usepackage{epstopdf}
\usepackage{wasysym}
\usepackage{bm}
\usepackage{graphicx}

\newenvironment{sciabstract}{%
\begin{quote} \bf}
{\end{quote}}

\newenvironment{methods}{%
    \section*{Methods}%
    \setlength{\parskip}{0pt}%
    }{}

\makeatletter
\def\@cite#1#2{\textsuperscript{{#1\if@tempswa , #2\fi}}}
\makeatother

\newcounter{lastnote}

\renewcommand{\figurename}{{\bf{Figure}}}

\makeatletter
\makeatletter \renewcommand{\fnum@figure}{{\bf{\figurename~\thefigure}}}
\makeatother

\title{Persistence of vortexlike phase fluctuations in underdoped to heavily overdoped Bi-2201 cuprates}

\author
{J. Terzic,$^{1,2}$ Bal K. Pokharel,$^{1,3, \dag}$ Z. Z. Li,$^4$, P. Senzier,$^4$ H. Raffy,$^{4}$ S. Ono,$^{5}$\\ 
Dragana Popovi\'{c}$^{1,3\ast}$\\
\\
\normalsize{$^{1}$National High Magnetic Field Laboratory, Florida State University,}\\
\normalsize{Tallahassee, Florida 32310, USA}\\
\normalsize{$^{2}$Department of Physics, Western Kentucky University,}\\
\normalsize{Bowling Green, Kentucky 42101, USA}\\
\normalsize{$^{3}$Department of Physics, Florida State University,}\\
\normalsize{Tallahassee, Florida 32306, USA}\\
\normalsize{$^{4}$Universit\'{e} Paris-Saclay, CNRS, Laboratoire de Physique des Solides,}\\
\normalsize{91405 Orsay, France}\\
\normalsize{$^{5}$International Center for Synchrotron Radiation Innovation Smart, Tohoku University,}\\
\normalsize{Aoba-Ku, Sendai 980-8572, Japan}\\
\\
\normalsize{$^{\dag}$ Present address: Intel, Hillsboro, Oregon 97124, USA}\\
\normalsize{$^\ast$To whom correspondence should be addressed; E-mail: dragana@magnet.fsu.edu}
}

\date{}

\begin{document} 


\baselineskip24pt


\maketitle 


\begin{sciabstract}
The mechanism that controls the superconducting (SC) transition temperature $\bm{T_{\mathrm{c}}^{0}}$ as a function of doping is one of the central 
questions in cuprate high-temperature superconductors.  While it is generally accepted that $\bm{T_{\mathrm{c}}^{0}}$ in underdoped cuprates is not determined by the scale of pairing but by the onset of global phase coherence, the role of phase fluctuations in the overdoped region has been controversial.  Here, our transport measurements in perpendicular magnetic fields ($\bm{H}$) on underdoped Bi-2201 reveal immeasurably small Hall response for $\bm{T>T_{\mathrm{c}}(H)}$ as a signature of SC phase with vortexlike phase fluctuations.  We find that the extent of such a regime in $\bm{T}$ and $\bm{H}$ is suppressed near optimal doping but becomes strongly enhanced in heavily overdoped Bi-2201.  Our results thus show that vortexlike phase fluctuations play an important role in the field-tuned SC transition in the heavily overdoped region, in contrast to conventional mean-field Bardeen-Cooper-Schrieffer description.  The unexpected nonmonotonic dependence of phase fluctuations on doping provides a new perspective on the SC transition in cuprates.
\end{sciabstract}

In copper oxides, superconducting state emerges by doping holes or electrons into an antiferromagnetic (AF) Mott insulator.  The doping is accomplished either by varying the oxygen concentration or by substitution with nonisovalent elements, suppressing the AF order of the parent compound, and giving rise to a dome-shaped $T_{\mathrm{c}}^{0}$ line with increasing concentration of charge carriers ($p$) in the CuO$_2$ planes.  In the underdoped region, where $T_{\mathrm{c}}^{0}(p)$ increases, the superfluid density is low due to the proximity to the Mott insulator\cite{Uemura1989,Emery1995,Lee2006}, implying a significant role of SC phase fluctuations.  The fluctuations are further enhanced by the effective two-dimensional (2D) nature of underdoped cuprates.   As a consequence, the thermodynamic SC transition occurs at $T_{\mathrm{c}}^{0}$ where long-range phase coherence is established, whereas the onset of SC pairing takes place at some higher, crossover temperature\cite{Keimer2015}.  This is in contrast to conventional superconductors, where the SC transition is well-described by the mean-field Bardeen-Cooper-Schrieffer (BCS) theory, according to which the SC phase coherence and the formation of Cooper pairs occur simultaneously.  Within that framework, fluctuations are negligible and, indeed, thermal SC fluctuations are observable within a small temperature range $\delta T$ such that $\delta T/T_{\mathrm{c}}^{0}\ll 1$ (ref.\cite{Larkin2009}).  On the other hand, the BCS theory was long-believed to provide an adequate description of the cuprate overdoped region, where the number of charge carriers is high and the anisotropy of the material is reduced.  However, this view has been challenged recently by several experiments performed at $H=0$ on heavily hole-doped cuprates\cite{Bozovic2016, Mahmood2019, He2021, Tromp2023}.  Although arguments have been put forward that experimental observations can still be explained  within the context of the BCS theory by properly taking into account the inevitable presence of disorder\cite{Lee2020, Pal2023} (``dirty $d$-wave'' theory), others have argued that disorder in a $d$-wave superconductor such as cuprates will lead to an emergent granular SC state\cite{Li2021-Kiv}, implying the importance of phase fluctuations beyond the mean-field treatment.  Real-space imaging in (Pb, Bi)$_2$Sr$_2$CuO$_{6+\delta}$ (Bi-2201) at $H=0$ has shown\cite{Tromp2023} that a state with SC grains embedded in a metallic matrix does indeed develop at high doping, but the results were inconsistent with the BCS theory for spatially heterogeneous systems and also seemed to rule out the effects of thermal phase fluctuations.  
Therefore, the role of both thermal and quantum phase fluctuations in the overdoped region remains an open question\cite{Sous2023}.

On the underdoped side, the existence of a broad phase fluctuation regime (at $T>T_{\mathrm{c}}^{0}$) was widely accepted, but its precise extent in temperature remained a subject of debate for many years\cite{Keimer2015}.  Likewise, in magnetic fields $H$ applied perpendicular to CuO$_2$ planes, the extent of SC phase with vortices, as well as its interplay with various charge and spin orders, were long controversial, especially at high $H$ as  $T\rightarrow 0$.  Those issues were essentially resolved in a series of linear and nonlinear transport experiments on La-214 cuprates\cite{Shi2014, Li2019, Shi2020}, which determined the in-plane $T$--$H$ vortex phase diagrams over an unprecedented range of $T$ and $H$.  It was shown that the vortex lattice with zero in-plane resistivity ($\rho_{\mathrm{xx}}=0$) is separated from the anomalous high-field normal state by a wide regime of a viscous vortex liquid (VL), i.e. SC phase fluctuations, that persist in 2D CuO$_2$ layers.  In this intermediate field range, the viscous VL freezes into a vortex glass at $T_\mathrm{c}(H)=0$.  Thus, at low $T\rightarrow 0$, increasing $H$ destroys superconductivity by quantum phase fluctuations\cite{Shi2014, Shi2020}.  At low fields, below the quantum melting field of the vortex lattice where $T_\mathrm{c}(H)\rightarrow0$, vortex liquid is observable at $T>T_{\mathrm{c}}(H)$ such that $\delta T(H)/T_{\mathrm{c}}(H)> 1$.  Here, including at $H=0$, thermal phase fluctuations lead to the destruction of superconductivity.  These results also demonstrated that the vortex phase diagrams of La-214 underdoped cuprates are qualitatively the same regardless of the type or strength of charge orders\cite{Shi2014, Li2019, Shi2020}, strongly suggesting that qualitatively the same vortex phase diagrams should apply to all underdoped cuprates.  Hence, similar experiments on another cuprate family, preferably one that does not exhibit spin order coexisting with superconductivity in contrast to underdoped La-214, are needed to verify that conclusion.  The most important question, though, is whether the VL regime persists to the heavily overdoped regions of the cuprate phase diagram where the (normal state) pseudogap closes.

Therefore, to address the above questions about the role of SC phase fluctuations, here we focus on doped Bi$_{2}$Sr$_2$CuO$_{6}$ (i.e. Bi-2201), which has several properties that make it an ideal system for this study.  Bi-2201 has a simple single-band Fermi surface and, using the chemical substitution or by varying the oxygen concentration, the parent compound can be easily doped all the way to the edge of the SC dome without undergoing Lifshitz transitions\cite{Ding2019}.  Just like La-214, Bi-2201 has a single CuO$_2$ layer per unit cell, and its relatively low $T_{\mathrm{c}}^{0}$ allows us to fully suppress superconductivity with $H$ and probe deep into the field-revealed normal state.  Bi-2201 has a relatively large residual resistivity\cite{Ono2003, Putzke2021, Berben2022, Juskus2024}, suggesting the importance of disorder, and in contrast to La-214, only short-range charge order has been reported\cite{Peng2016, Peng2018, Li2021}, with no evidence of spin order\cite{Peng2018_2}.  We combine linear in-plane resistivity $\rho_{\mathrm{xx}}$ and Hall resistivity $\rho_{\mathrm{yx}}$ with nonlinear in-plane transport or voltage--current ($V$--$I$) characteristics to probe both charge and vortex matter over a wide range of $10^{-2}\lesssim T/T_{\mathrm{c}}^{0} \lesssim 10$ down to mK temperatures and fields up to $H/T_{\mathrm{c}}^{0}\gtrsim 10$~T/K for several values of doping.  In the underdoped region, we find that, qualitatively, the $T$--$H$ vortex phase diagram is the same as that in the La-214 family\cite{Shi2014, Li2019, Shi2020} and, thus, independent of the presence of spin order.  This result establishes the universality of the vortex phase diagram in underdoped cuprates.  In addition, we reveal a vanishing Hall response ($\rho_{\mathrm{yx}}=0$) as a signature of the (viscous) VL regime with $\rho_{\mathrm{xx}}\neq 0$.  By tracking the evolution of such a regime with doping, we obtain our key results.  First, the VL does persist in the heavily overdoped region, such that it is observable up to $\delta T(H)/T_{\mathrm{c}}(H)> 1$.   Therefore, complementing previous $H=0$ studies\cite{Bozovic2016, Mahmood2019, He2021, Tromp2023} that used other techniques, our magnetotransport measurements indicate that the SC transition in heavily overdoped cuprates is of a non-mean-field type.  Second, in contrast to a commonly held belief, the doping dependence of 
$\delta T(H)/T_{\mathrm{c}}(H)$ is nonmonotonic: phase fluctuations are considerably suppressed near optimal doping compared to both underdoped and heavily overdoped regions, resulting in a behavior more reminiscent of a BCS-type superconductor.\\
\vspace{-12pt}

\begin{figure}
\centering
\includegraphics[width=0.92\textwidth,clip=]{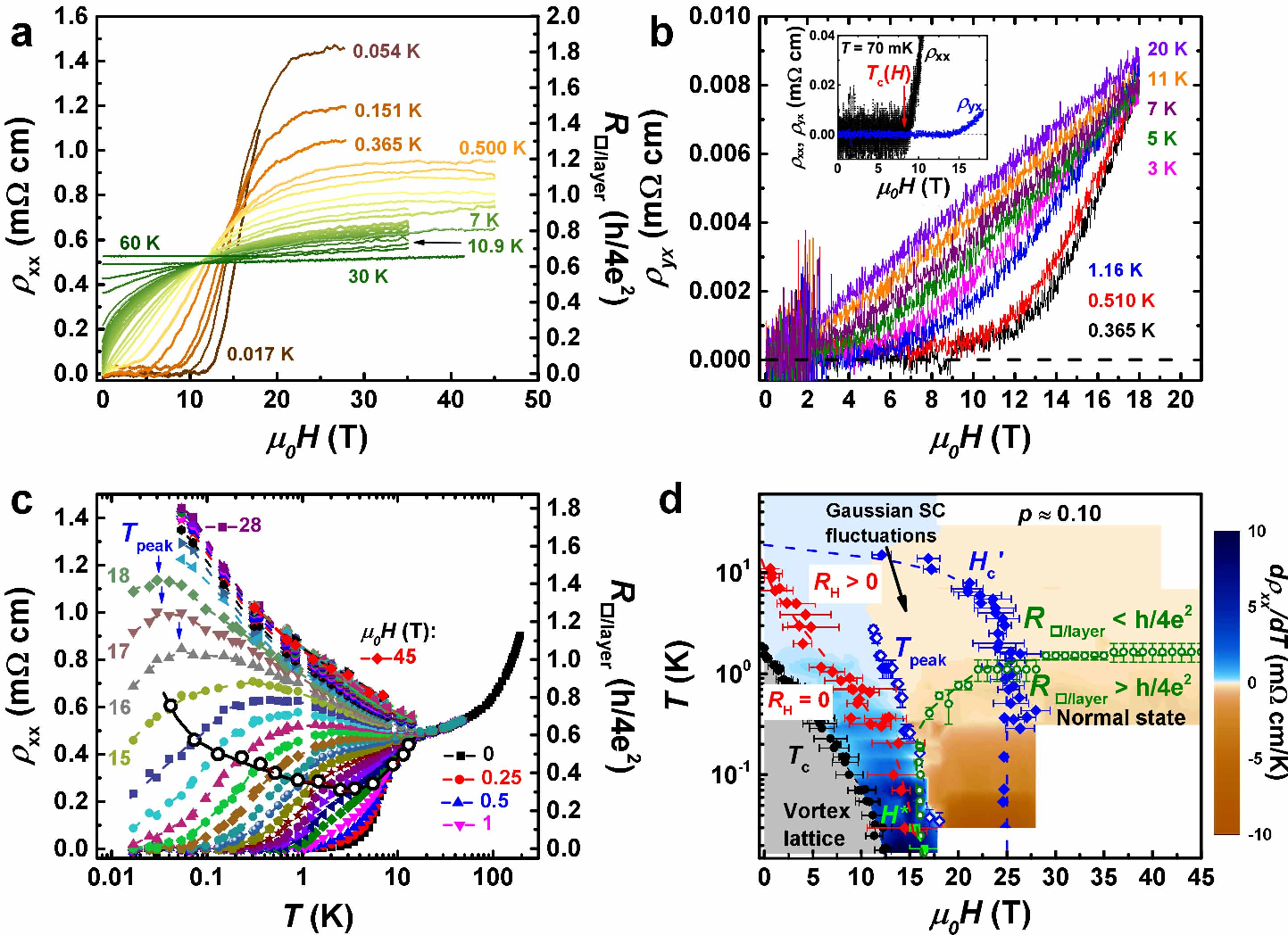}
\caption{\textbf{In-plane magnetotransport in underdoped Bi-2201 ($\bm{p \approx 0.10}$).}  \textbf{a} $\rho_\mathrm{xx}$ vs $H \parallel c$ at several $T$, as shown.  Right axis: the corresponding $R_{\square/\mathrm{layer}}$ in units of quantum resistance for Cooper pairs, $R_\mathrm{Q}=h/4e^2$. \textbf{b} $\rho_\mathrm{yx}$ vs $H$ up to 18~T at several $T$, as shown. Inset: Difference in the onsets of nonzero $\rho_\mathrm{yx}$ (blue curve) and $\rho_\mathrm{xx}$ (black curve) at $T= 0.07$~K; red arrow indicates $T_{\mathrm{c}} (H)$, i.e. the onset of nonzero $\rho_\mathrm{xx}$.  \textbf{c} $\rho_\mathrm{xx}(T)$ for several $H$ up to 45~T; the data for $1\leq H$(T)$\leq 45$ are shown in steps of 1~T.  Blue arrows indicate the peak in $\rho_\mathrm{xx}(T)$, $T_{\mathrm{peak}}(H)$; right axis: the corresponding $R_{\square/\mathrm{layer}}$.  Open black circles: $T_{R_\mathrm{H} =0}(H)$, onset of zero Hall coefficient $R_\mathrm{H} = 0$, for $0.25\leq H$[T]~$\leq 15$; solid black line guides the eye. \textbf{d} $T$--$H$ phase diagram  
($H\parallel c$); color map: $d\rho_{\mathrm{xx}}/dT$. $T_{\mathrm{c}} (H)$ (black dots): boundary of the pinned vortex lattice in which $\rho_\mathrm{xx} (T < T_{\mathrm{c}} )= 0$ and $\rho_\mathrm{yx}=0$, as expected for a superconductor.  The quantum melting field of the vortex lattice where $T_{\mathrm{c}} (H)\rightarrow 0$ is $\sim 12.8$~T.  Open blue diamonds: $T_{\mathrm{peak}}(H)$; solid blue diamonds: $H_{\mathrm{c}}' (T)$, the fields above which Gaussian SC fluctuations are not observed.  The error bars for $H_{\mathrm{c}}' (T)$ correspond to $\pm 1$~SD (standard deviation) in the slopes of the linear fits in Supplementary Fig.~1a. The dashed line is a fit with $\mu_0 H_{\mathrm{c}}'$~[T] $= (25\pm 2)[1-(T [\mathrm{K}]/(18.9\pm 0.9)^2]$.  $H^{\ast}(T)$ (neon squares):  boundary between the viscous vortex liquid with non-Ohmic $dV/dI$ for $H<H^{\ast}(T)$ and Ohmic behavior at 
$H>H^{\ast}(T)$; error bars reflect the uncertainty in determining $H^{\ast}(T)$ within experimental resolution.  Open green circles separate the low-$T$, high-$H$ regime where $R_{{\square}/\mathrm{layer}} > R_Q$ from the regime where $R_{{\square}/\mathrm{layer}} < R_Q$.  Red diamonds: $T_{R_\mathrm{H} =0}(H)$, boundary between the regime with $R_\mathrm{H}>0$, found at higher $H$ and $T$, and the $R_\mathrm{H}=0$ regime; the dashed red line guides the eye. 
}
\label{fig:under}
\end{figure}

\noindent\textbf{Underdoped region}\\ 
Underdoped Bi-2201 samples are single crystals of Bi$_2$Sr$_{2-x}$La$_x$CuO$_{6+\delta}$ with $x = 0.84$ and $p \approx 0.10$ (see Methods).   Similar to earlier measurements\cite{Ono2000} of the in-plane resistivity down to $T=0.7$~K, our study down to much lower $\sim 0.05$~K finds that $\rho_{\mathrm{xx}}(H)$ at low enough $T$ increases (Fig.~\ref{fig:under}a), indicating a suppression of superconductivity by $H$.  This is followed by a 
%
%
weak field dependence of $\rho_\mathrm{xx}$ at higher $H$, representing the normal-state behavior.  The $\rho_{\mathrm{xx}}(H)$ data were used to determine $T_{\mathrm{c}}(H)$, the melting temperature of the vortex lattice in which $\rho_{\mathrm{xx}}=0$, as well as $H_{\mathrm{c}}' (T)$, the onset of Gaussian fluctuations of the SC amplitude and phase.  As usual\cite{Rullier2007, Rourke2011, Shi2014, Shi2020}, $H_{\mathrm{c}}' (T)$ is defined as the field above which the magnetoresistance increases as $H^2$, as expected in the high-$T$ normal state (Supplementary Fig.~1).   

Figure~\ref{fig:under}b shows $\rho_{\mathrm{yx}}(H)$ at several $T$ in fields up to 18~T.  (Additional results, including those obtained up to 41~T, are presented in Supplementary Figs.~2a-d; see also Supplementary Figs.~3a-d for the corresponding Hall coefficient $R_{\mathrm{H}}=\rho_{\mathrm{yx}}(H)/H$, as well as Supplementary Fig.~4 for $\rho_{\mathrm{yx}}(H)$ and $R_{\mathrm{H}}(H)$ at various $T$.)  At the highest $T$, $\rho_\mathrm{yx}\propto H$, as expected in conventional metals.  At lower $T$, as the magnetic field increases and suppresses superconductivity, the Hall signal increases and then recovers its conventional $\rho_\mathrm{yx}\propto H$ behavior at high $H$ (Supplementary Figs.~2a-d), i.e. $R_{H}$ becomes independent of the field (Supplementary Figs.~3a-d; also Supplementary Fig.~4).  The results are in good agreement with previous Hall measurements in underdoped Bi-2201, which focused on exploring the Hall coefficient in the normal state\cite{Balakirev2003}.  Here, in contrast, we focus on the low-field regime, where $\rho_\mathrm{yx}$ starts to deviate from zero.  Surprisingly, we observe a large, up to several teslas, difference in the onset fields of nonzero $\rho_\mathrm{xx}$, which defines $T_\mathrm{c}$ for that field, and nonzero $\rho_\mathrm{yx}$ (Fig.~\ref{fig:under}b inset).  The onsets of $\rho_\mathrm{yx}=0$, i.e. $R_{\mathrm{H}}=0$, are also shown in the $\rho_\mathrm{xx}(T)$ data for various fields (Fig.~\ref{fig:under}c) to help determine the extent of the $R_{\mathrm{H}}=0$ regime in $T$ above $T_\mathrm{c}(H)$.  Following their initial suppression by increasing $H$, the onsets of the $R_{\mathrm{H}}=0$ regime appear to track $T_{\mathrm{peak}}(H)$, the position of the peak in $\rho_{\mathrm{xx}}(T)$, which emerges at intermediate fields.  In underdoped La-214 cuprates, $T_{\mathrm{peak}}(H)$ was identified\cite{Shi2014, Shi2020} as the onset of the viscous VL observed at $T<T_{\mathrm{peak}}(H)$.

Figure~\ref{fig:under}d summarizes our findings on underdoped Bi-2201 in a $T$--$H$ phase diagram established across a large range of $T$ and $H$, and showing the evolution of $T_{\mathrm{c}} (H)$,  $H_{\mathrm{c}}' (T)$, $T_{\mathrm{peak}}(H)$, as well as the onsets of $R_\mathrm{H}=0$ (see also  Supplementary Figs.~3a, b for the determination of $R_\mathrm{H}=0$).  It is obvious that there is a fairly wide regime with $\rho_\mathrm{xx}\neq 0$ and $R_\mathrm{H}=0$ observed at $T>T_{\mathrm{c}}$ for a given $H$, which is distinct from the pinned vortex lattice characterized by $\rho_\mathrm{xx}(T<T_{\mathrm{c}})= 0$ and $R_\mathrm{H}=0$.  At low $T$, the boundary of the $R_\mathrm{H}=0$ regime indeed approaches $T_{\mathrm{peak}}(H)$.  

To confirm the existence of the viscous VL at intermediate fields, between $T_{\mathrm{c}} (H)$ and $T_{\mathrm{peak}}(H)$, 
 we use two methods.  First, we consider the $T$-dependence of the linear ($I_{\mathrm{dc}}\rightarrow 0$) resistivity $\rho_{\mathrm{xx}}(T)$.  We establish that, for $T<T_{\mathrm{peak}}(H)$, it is described best with the power-law fits $\rho_{\mathrm{xx}} (H,T) = k(H)T^{\alpha(H)}$, where $\alpha(H)$ decreases to zero with $H$ (Supplementary Fig.~5a, b).  The power-law dependence suggests that, at intermediate fields, a true SC state ($\rho_\mathrm{xx}= 0$) exists only at $T = 0$ when the vortices are frozen.  Similar behavior was found\cite{Shi2014, Shi2020} in underdoped La-214 cuprates, consistent with the expectations for a viscous VL above its glass freezing temperature $T_{\mathrm{g}} = 0$.  Second, we measure nonlinear transport, i.e. differential resistance $dV/dI$ as a function of the dc current excitation $I_{\mathrm{dc}}$ (Methods).  In the same field range at low $T$, we find that $dV/dI$ is non-Ohmic for $I_{\mathrm{dc}}\neq 0$, although the linear resistance ($dV/dI$ for $I_{\mathrm{dc}}\rightarrow 0$) is not zero (Supplementary Fig.~5c, d).  This type of behavior is precisely what is expected from the motion of vortices in the presence of disorder, i.e. it is a signature of a viscous VL\cite{Doussal2010}.   At higher $H$, Ohmic behavior is recovered (Supplementary Fig.~5d), allowing us to determine fields $H^{\ast}(T)$ that separate the viscous VL regime at lower $H$ from the Ohmic regime characteristic of the normal state at high $H$ (Fig.~\ref{fig:under}d).  Just like in underdoped La-214 cuprates\cite{Shi2014, Shi2020}, as $T\rightarrow 0$ there is a quantitative agreement between $H^{\ast}(T)$, the boundary of the viscous VL obtained from nonlinear transport, with the values of $T_{\mathrm{peak}}(H)$ obtained from the linear resistivity measurements.  Moreover, here these two characteristic energy scales, which indicate the vanishing of the SC phase with vortices, also follow  the ``$h/4e^2$'' line (Fig.~\ref{fig:under}d), where the sheet resistance $R_{\square/\mathrm{layer}}\approx R_Q$, with $R_\mathrm{Q}=h/(2e)^2$ the quantum resistance for Cooper pairs.   This suggests the onset of localization of Cooper pairs in CuO$_2$ planes\cite{MPAF1990}, consistent with the observed crossover at $T_{\mathrm{peak}}(H)$ to the well-known, albeit little-understood anomalous normal state (Fig.~\ref{fig:under}d) with a weak, insulating $\rho_{xx}\propto\ln (1/T)$ behavior\cite{Ando1995, Ono2000} for $H>H_{\mathrm{c}}'$ (Fig.~\ref{fig:under}c).  Therefore, by using complementary transport techniques that are sensitive to global phase coherence, we have established that the regime with $R_\mathrm{H}=0$  and $\rho_\mathrm{xx}\neq 0$ clearly represents a signature of the SC phase with vortexlike phase fluctuations.  

Our results demonstrate that the vortex phase diagram in underdoped Bi-2201 ($p \approx 0.10$) is qualitatively the same as that in the La-214 family and, thus, independent of the details of either spin or charge orders.  This universality observed in the underdoped region is further supported by $\rho_\mathrm{xx}$ vs $I$ measurements\cite{Hsu2021} in Bi-2201 with $p \approx 0.125$ for  $\mu_{0}H/T^{0}_{c}\approx 1.2$ and YBa$_2$Cu$_4$O$_8$ with $p= 0.14$ for  $\mu_{0}H/T^{0}_{c}\approx 0.6$, which provided evidence for the presence of a vortex liquid.  Furthermore, we show that Hall effect measurements and, in particular, the regime with $R_\mathrm{H}=0$ and $\rho_\mathrm{xx}\neq 0$, can be used to identify the presence of a nearly frozen or pinned vortex liquid.  Therefore, we focus on this method, employed along other transport techniques, to determine vortex phase diagrams in Bi-2201 at higher doping, all the way to the heavily overdoped region.  \\
\vspace*{-12pt}

\begin{figure}
\centerline{\includegraphics[width=\textwidth]{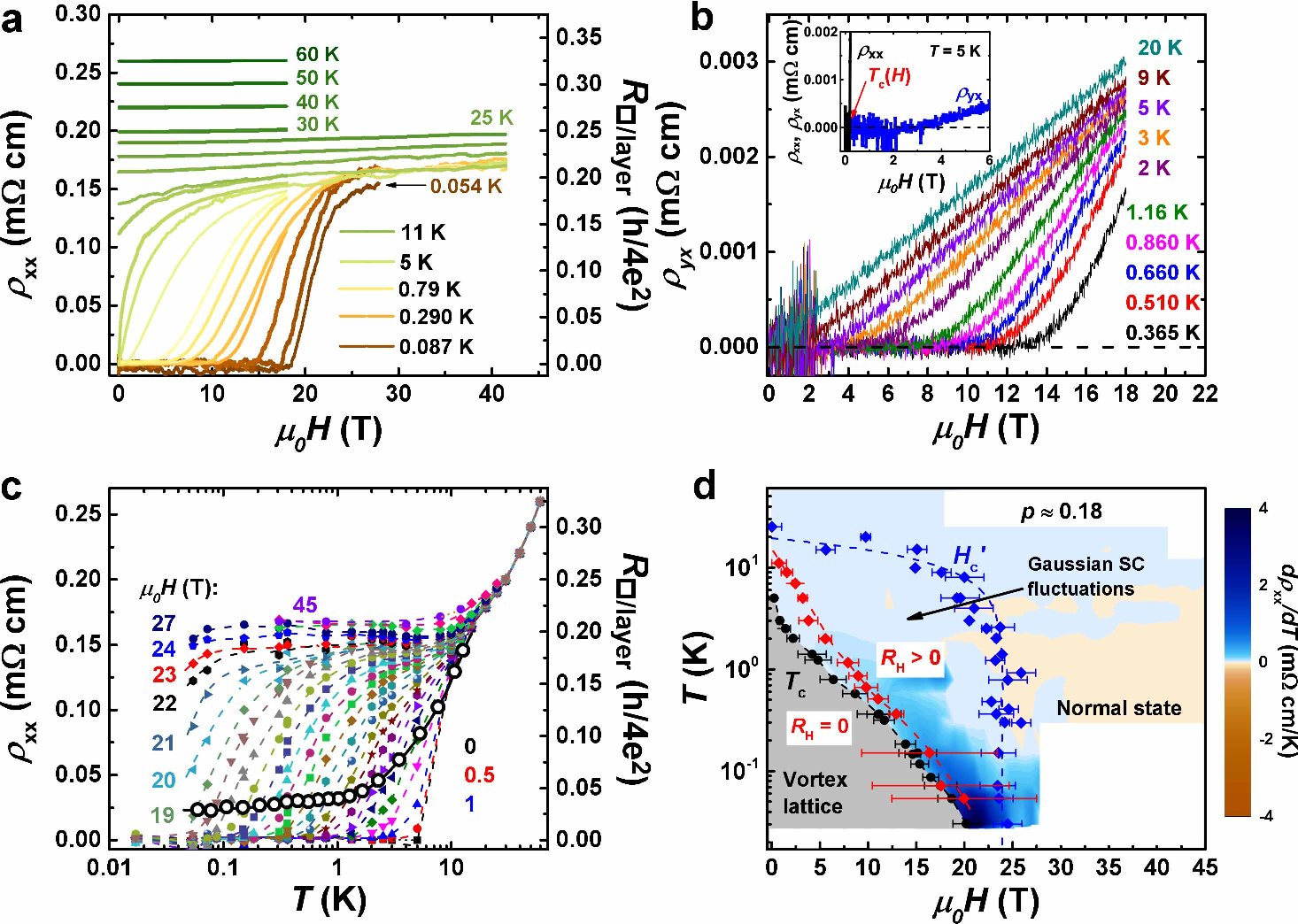}}
\caption{\textbf{In-plane magnetotransport in weakly overdoped Bi-2201 ($\bm{p \approx 0.18}$).}  \textbf{a} $\rho_\mathrm{xx}$ vs $H \parallel c$ at several $T$, as shown.  Right axis: the corresponding $R_{\square/\mathrm{layer}}$ in units of quantum resistance for Cooper pairs, $R_\mathrm{Q}=h/4e^2$. \textbf{b} $\rho_\mathrm{yx}$ vs $H$ up to 18~T at several $T$, as shown. Inset: Difference in the onsets of nonzero $\rho_\mathrm{yx}$ (blue curve) and $\rho_\mathrm{xx}$ (black curve) at $T= 5$~K; red arrow indicates $T_{\mathrm{c}} (H)$, i.e. the onset of nonzero $\rho_\mathrm{xx}$. \textbf{c} $\rho_\mathrm{xx}(T)$ for several $H$ up to 45~T, as shown; the data for $1\leq H$(T)$\leq 19$ are given in steps of 1~T, while the highest $H$ curves correspond to 45~T, 38~T, and 35~T, respectively.  The right axis shows the corresponding $R_{\square/\mathrm{layer}}$.  Open black circles: $T_{R_\mathrm{H} =0}(H)$, onset of zero Hall coefficient $R_\mathrm{H} = 0$, for $0.5\leq H$[T]~$\leq 19$; solid black line guides the eye. \textbf{d} $T$--$H$ phase diagram ($H\parallel c$); color map: $d\rho_{\mathrm{xx}}/dT$. $T_{\mathrm{c}} (H)$ (black dots): boundary of the pinned vortex lattice, in which $\rho_\mathrm{xx} (T < T_{\mathrm{c}} )= 0$ and $\rho_\mathrm{yx}=0$, as expected for a superconductor.  Solid blue diamonds: $H_{\mathrm{c}}' (T)$, the fields above which Gaussian SC fluctuations are not observed.  The error bars for $H_{\mathrm{c}}' (T)$ correspond to $\pm 1$~SD in the slopes of the linear fits in Supplementary Fig.~1b. The dashed line is a fit with $\mu_0 H_{\mathrm{c}}' $~[T] $= (24\pm 3)[1-(T [\mathrm{K}]/(19\pm 2)^2]$.  Red diamonds: $T_{R_\mathrm{H} =0}(H)$, boundary between the regime with $R_\mathrm{H}>$ 0, found at higher $H$ and $T$, and the 
$R_\mathrm{H}= 0$ regime; the dashed red line guides the eye.
}
\label{fig:over}
\end{figure}

\noindent\textbf{Weakly overdoped region}\\
Weakly overdoped Bi-2201 samples are single crystals of Bi$_{2.1}$Sr$_{1.9}$CuO$_{6+\delta}$ with hole concentration ${p} \approx 0.18$ (see Methods).   As in the underdoped region, we measure $\rho_{\mathrm{xx}}(H)$ and $\rho_{\mathrm{yx}}(H)$ at different $T$ [Fig.~\ref{fig:over}a and Fig.~\ref{fig:over}b, respectively; see also Supplementary Figs.~2e-h for additional $\rho_{\mathrm{yx}}(H)$ data, i.e.  Supplementary Figs.~3e-h for the corresponding $R_{\mathrm{H}}(H)$].
%
%
Here we also observe a clear difference in the onset fields of nonzero $\rho_\mathrm{xx}$ and nonzero $\rho_\mathrm{yx}$ (Fig.~\ref{fig:over}b inset).  The onsets of $R_{\mathrm{H}}=0$ are shown in the $\rho_\mathrm{xx}(T)$ data for various fields (Fig.~\ref{fig:over}c), as well as in the $T$--$H$ phase diagram (Fig.~\ref{fig:over}d).  

It is apparent that the regime with $R_\mathrm{H}=0$ and $\rho_\mathrm{xx}\neq 0$, indicative of the nearly frozen or pinned VL, is substantially suppressed compared to the underdoped region (cf. Fig.~\ref{fig:under}c, d), in agreement with general expectations\cite{Emery1995}.  In particular, here 
$R_\mathrm{H}=0$ with $\rho_\mathrm{xx}\neq 0$ is observed only at lower $H\lesssim 20$~T, i.e., for $T> T_\mathrm{c}(H)$ (Fig.~\ref{fig:over}d).  In this range of fields, $\rho_\mathrm{xx}(T)$ (Fig.~\ref{fig:over}c; also Supplementary Fig.~6) is indeed reminiscent of the behavior of a BCS-type superconductor, indicating relatively weak thermal SC fluctuations.  At intermediate fields $20\lesssim H$(T)$\lesssim H_{\mathrm{c}}'(T=0)=24$ where Gaussian SC fluctuations are still observed (Fig.~\ref{fig:over}d), a peak in $\rho_\mathrm{xx}(T)$ is not found at low $T$ (Fig.~\ref{fig:over}c).  Instead, at these fields at the lowest $T<0.1$~K, $\rho_\mathrm{xx}$ seems to drop with decreasing $T$ in a manner similar to that at low fields for $T> T_\mathrm{c}(H)$ (Supplementary Fig.~6), in contrast to the behavior in the underdoped region (Supplementary Fig.~5a, b).  In addition, nonlinear $dV$/$dI$ was not observed either, suggesting the absence of a viscous VL at intermediate fields.  Current-dependent $\rho_\mathrm{xx}$ was also not observed\cite{Hsu2021} in Bi$_2$Sr$_{2-x}$La$_x$CuO$_{6+\delta}$ with $p \approx 0.19$ for $\mu_{0}H/T^{0}_{c}\approx 1.3$.  Finally, at the highest fields [$H>H_{\mathrm{c}}'(T=0)=24$~T] where SC fluctuations are suppressed, $\rho_\mathrm{xx}(T)$ is negligible, indicative of the metallic normal state, in agreement with the Hall effect measurements (Supplementary Fig.~7) and previous work\cite{Ono2000}.  
\\
\vspace*{-12pt}

\begin{figure}
\centerline{\includegraphics[width=\textwidth]{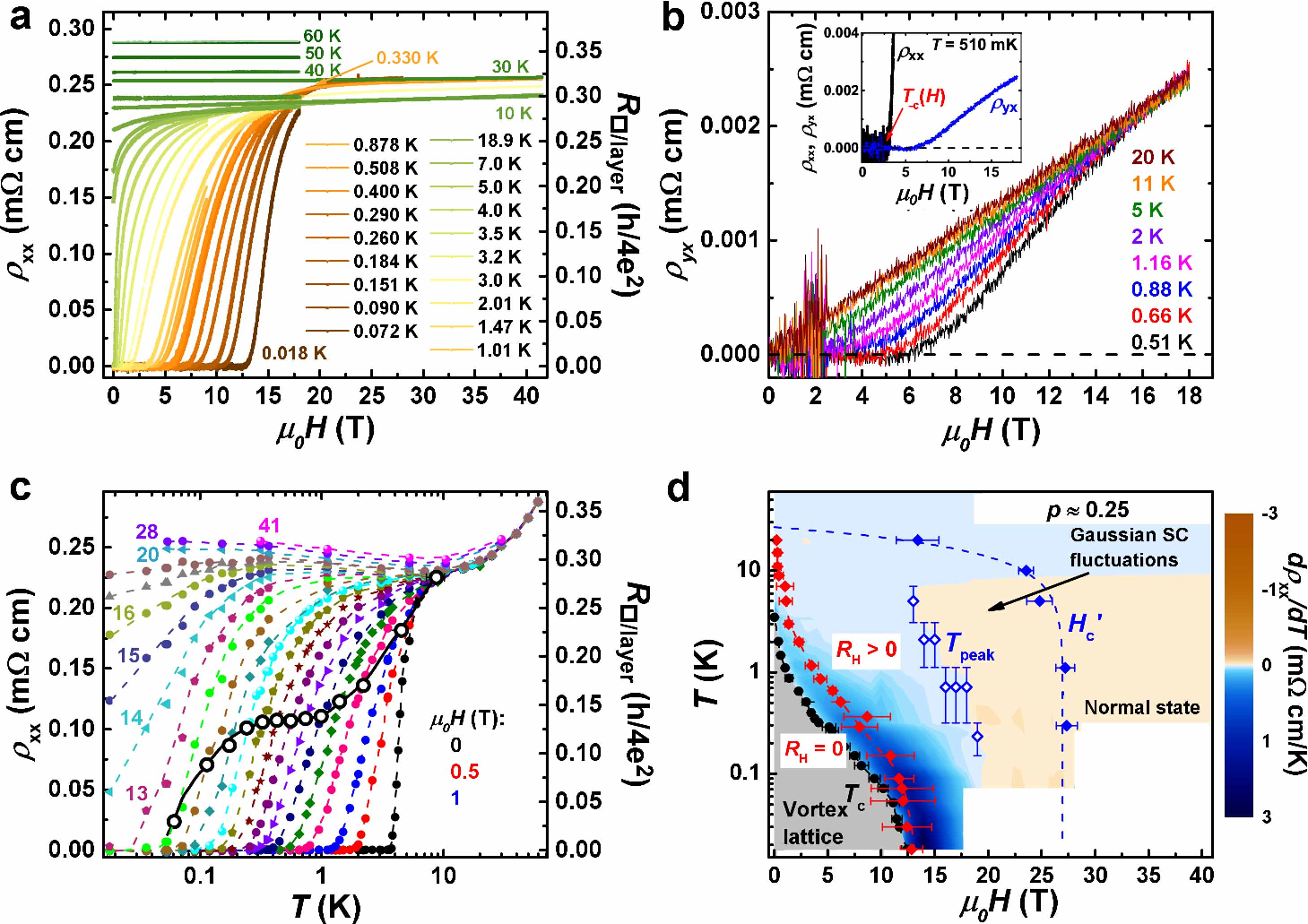}}
\caption{\textbf{In-plane magnetotransport in heavily overdoped Bi-2201 ($\bm{p \approx 0.25}$).}  \textbf{a} The in-plane longitudinal resistivity $\rho_\mathrm{xx}$ vs $H\parallel c$ at several $T$, as shown. The right axis shows the corresponding $R_{\square/\mathrm{layer}}$ in units of quantum resistance for Cooper pairs, $R_\mathrm{Q}=h/4e^2$. \textbf{b} Hall resistivity $\rho_\mathrm{yx}$ vs $H$ up to 18~T at several $T$, as shown. Inset: Difference in the onsets of nonzero $\rho_\mathrm{yx}$ (blue curve) and $\rho_\mathrm{xx}$ (black curve) at $T = 0.510$~K; red arrow indicates $T_{\mathrm{c}}(H)$, i.e. the onset of nonzero $\rho_\mathrm{xx}$. \textbf{c} $\rho_\mathrm{xx}(T)$ for several $H$ up to 41~T, as shown; the data for $1\leq H$(T)$\leq 18$ are shown in steps of 1~T.  The right axis shows the corresponding $R_{\square /\mathrm{layer}}$.  Open black circles indicate $T_{R_\mathrm{H} =0}(H)$, the onset of zero Hall coefficient $R_\mathrm{H }= 0$, for $0.5\leq H$[T]~$\leq 12$; solid black line guides the eye. \textbf{d} $T$--$H$ phase diagram ($H \parallel c$); color map: $d\rho_{\mathrm{xx}}/dT$. $T_{\mathrm{c}} (H)$ (black dots) mark the boundary of the pinned vortex lattice, which is a superconductor with $\rho_\mathrm{xx} = 0$ for all $T < T_{\mathrm{c}} (H)$.  Open blue diamonds: $T_{\mathrm{peak}}(H)$; solid blue diamonds: $H_{\mathrm{c}}' (T)$ , the fields above which Gaussian superconducting fluctuations are not observed.  The error bars for $H_{\mathrm{c}}' (T)$ correspond to $\pm~1$~SD in the slopes of the linear fits in Supplementary Fig.~1c. The dashed line is a fit with $\mu_0 H_{\mathrm{c}}' $~[T] $= (27\pm 1)[1-(T [\mathrm{K}]/(26.9\pm 0.7)^2]$. Red diamonds: $T_{R_\mathrm{H} =0}(H)$, the boundary between the region with $R_\mathrm{H}(T)>$ 0 at higher $H$ and $R_\mathrm{H}(T) = 0$ at lower $H$.  The latter generally extends above $T_\mathrm{c}(H)$.  The dashed red line guides the eye.
}
\label{fig:film}
\end{figure}

\noindent\textbf{Heavily overdoped region}\\
Heavily overdoped Bi-2201 samples are thin films of Bi$_{2}$Sr$_{2}$CuO$_{6+\delta}$ (see Methods).  Here $\rho_{\mathrm{xx}}(H)$ and $\rho_{\mathrm{yx}}(H)$ measurements at different $T$ on a film with the hole concentration ${p} \approx 0.25$ [Fig.~\ref{fig:film}a and Fig.~\ref{fig:film}b, respectively; see also Supplementary Figs.~2i-l for additional $\rho_{\mathrm{yx}}(H)$ data, i.e.  Supplementary Figs.~3i-l for the corresponding $R_{\mathrm{H}}(H)$]
%
%
also reveal a large difference in the onset fields of nonzero $\rho_\mathrm{xx}$ and nonzero $\rho_\mathrm{yx}$ (Fig.~\ref{fig:film}b inset), i.e., a regime with $R_\mathrm{H}=0$ and $\rho_\mathrm{xx}\neq 0$ observed over a wide range of $T > T_{\mathrm{c}} (H)$ and $H<13$~T (Fig.~\ref{fig:film}c, d).  
Surprisingly, the suppression of this regime by $H$ (black circles in Fig.~\ref{fig:film}c) seems to occur less rapidly than in the weakly overdoped region (cf. black circles in Fig.~\ref{fig:over}c).  

At intermediate fields (13~T~$\lesssim H \lesssim 19$~T), a peak in $\rho_{\mathrm{xx}}(T)$ emerges (Fig.~\ref{fig:film}c).  Although not as pronounced as in the underdoped region (Fig.~\ref{fig:under}c), here we also find a power-law dependence characteristic of a viscous VL: $\rho_{\mathrm{xx}} (H,T) = k(H)T^{\alpha(H)}$ for $T<T_{\mathrm{peak}}(H)$, with $\alpha(H)$ decreasing to zero with $H$ (Supplementary Fig.~8).  While any nonlinear $dV/dI$ was too weak to be observed in this sample, we find that the peak in $\rho_{\mathrm{xx}}(T)$ becomes more pronounced with further increase in doping (see Supplementary Fig.~9 for a film with $p\approx 0.27$), thus confirming the presence of the intermediate viscous VL regime in heavily overdoped samples, outside of the pseudogap regime.  

At the highest fields  $H>H_{\mathrm{c}}'(T=0)\sim27$~T where SC fluctuations are fully suppressed (Fig.~\ref{fig:under}d), the normal-state $\rho_\mathrm{xx}(T)$ is weak, indicative of a metallic ground state, with $d\rho_{\mathrm{xx}}/dT<0$ suggesting the effects of disorder.  These findings are also in agreement with our Hall effect measurements (Supplementary Fig.~10).  Moreover, we note that the high-field, normal-state value of $R_\mathrm{H}$ obtained on the $p \approx 0.25$ Bi$_{2}$Sr$_{2}$CuO$_{6+\delta}$ film for $T\rightarrow 0$ (Supplementary Fig.~10b) is comparable to that found\cite{Putzke2021} at higher $T=(15-54)$~K on a Bi$_{2-y+x}$Pb$_y$Sr$_{2-x}$CuO$_{6+\delta}$ single crystal with a similar $p \approx 0.245$, thus demonstrating consistent behavior between films and single crystals.
\\
\vspace{-12pt}

\begin{figure}[!tb]
  \centering
     \includegraphics[width=\textwidth]{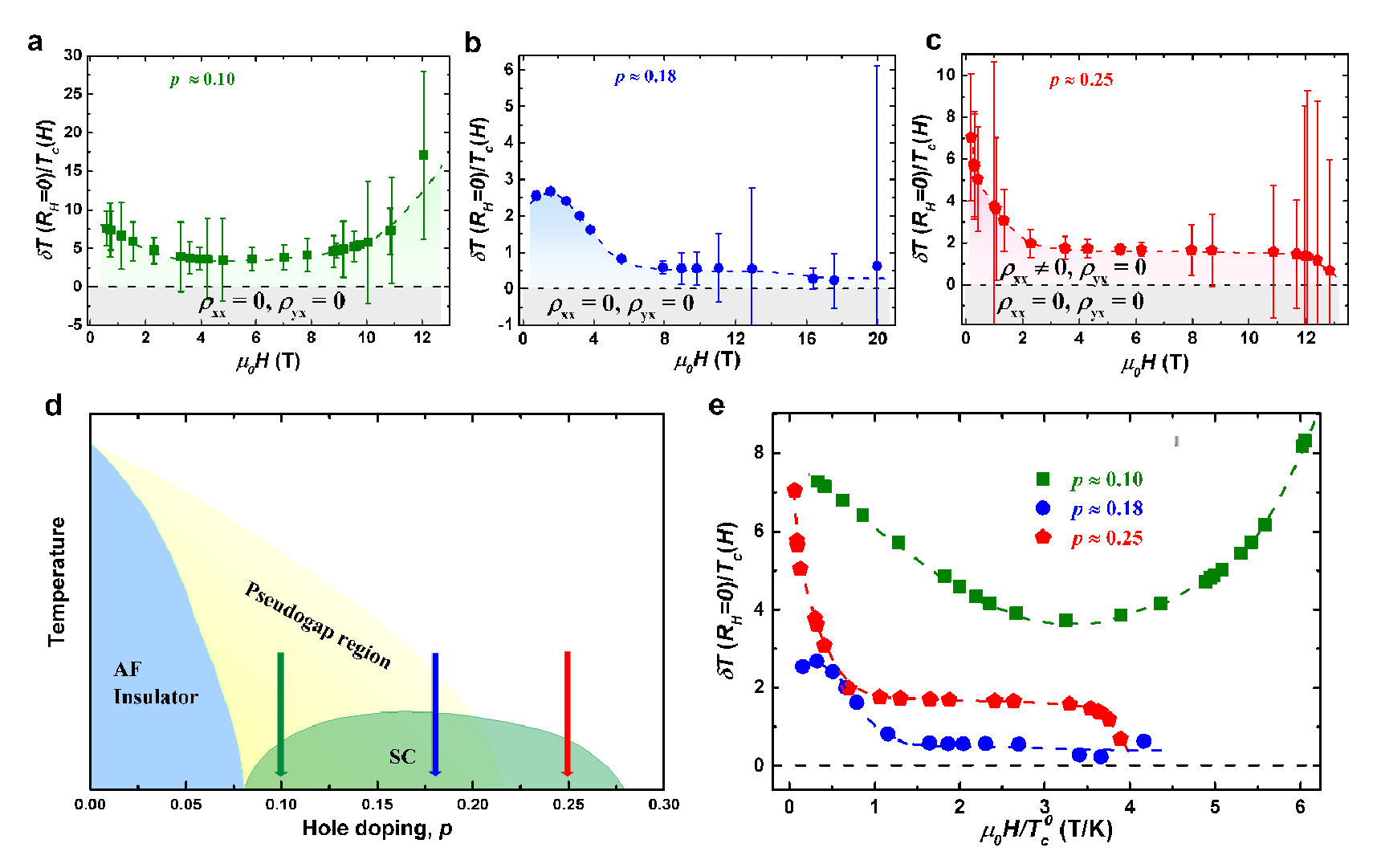}
\caption{\textbf{The extent of the regime with zero Hall coefficient in different doping regions.}  The normalized width of the $R_\mathrm{H} = 0$ regime above $T_\mathrm{c}(H)$, i.e. $\delta T(R_\mathrm{H} =0)/T_\mathrm{c}(H)=[T_{R_\mathrm{H} =0}(H)-T_{\mathrm{c}}(H)]/T_{\mathrm{c}}(H)$, vs magnetic field ($H\parallel c$) for \textbf{a,} $p \approx 0.10$, \textbf{b,} $p \approx 0.18$, and \textbf{c,} $p \approx 0.25$.  $\delta T(R_\mathrm{H} =0)$ were calculated from the fits of $T_{\mathrm{c}}(H)$ and $T_{R_\mathrm{H} =0}(H)$ to an exponential form, respectively, and the error bars for $\delta T(R_\mathrm{H} =0)/T_{\mathrm{c}}(H)$ result from $\pm 1$~SD (standard deviation) in those fits.  The areas below the dashed horizontal line at which $\delta T=0$ correspond to the pinned vortex lattice where both $\rho_{\mathrm{xx}}$ and $\rho_{\mathrm{yx}}$ are zero. The lightly shaded areas in \textbf{a}, \textbf{b}, and \textbf{c} are the $T_\mathrm{c}(H)<T< T_{R_\mathrm{H} =0}(H)$ regimes where $\rho_{\mathrm{xx}}$ is not zero, but $\rho_{\mathrm{yx}}$ remains immeasurably low; dashed lines guide the eye.  \textbf{d,}  The arrows show the doping values from \textbf{a}, \textbf{b}, and \textbf{c} in the $T$--$p$ phase diagram of the Bi-2201 family for easier comparison.  \textbf{e}, $\delta T(R_\mathrm{H} = 0)/T_\mathrm{c}(H)$ from \textbf{a}, \textbf{b}, and \textbf{c} plotted vs $\mu_{0}H/T^{0}_{c}$, where $T^{0}_{c}$ are the respective zero-field superconducting transition temperatures.  The error bars are omitted for clarity, and dashed lines guide the eye. 
 }
\label{Hall-all}
\end{figure}

\noindent\textbf{Evolution of vortexlike phase fluctuations with doping}\\
To examine the dependence of the $R_\mathrm{H}=0$, $\rho_\mathrm{xx}\neq 0$ regime on doping, we analyze its extent in $T$ above $T_{\mathrm{c}} (H)$.  In particular, we determine the relative width $\delta T(R_\mathrm{H} =0)/T_\mathrm{c}(H)=[T_{R_\mathrm{H} =0}(H)-T_{\mathrm{c}}(H)]/T_{\mathrm{c}}(H)$ for fixed $H$ for the values of $p$ that we have studied in detail (Fig.~\ref{Hall-all}a, b, c) in different doping regions of the $T$--$p$ phase diagram 
%
(Fig.~\ref{Hall-all}d).  In the underdoped region at low $H$ (Fig.~\ref{Hall-all}a), the relative width of the nearly frozen or pinned VL, as determined from $R_\mathrm{H}=0$ with $\rho_\mathrm{xx}\neq 0$, extends even up to $\delta T(H)/T_{\mathrm{c}}(H)\sim 7$ (Fig.~\ref{Hall-all}a), consistent with the findings in underdoped La-214 cuprates using different probes of vortex matter (refs.\cite{Shi2014, Shi2020, ZShi-PDW} and refs. therein).  Following its initial suppression by $H$, $\delta T(H)/T_{\mathrm{c}}(H)$ increases to comparably large values [$\delta T(H)/T_{\mathrm{c}}(H)\sim 8$] at high fields, reflecting the presence of the viscous VL at $T<T_{\mathrm{peak}}(H)$ (see Fig.~\ref{fig:under}c, d).   We recall that the viscous VL with $T_{\mathrm{c}}(H)=0$ persists to even higher $H\approx H^{\ast}$ (Fig.~\ref{fig:under}d), not shown in Fig.~\ref{Hall-all}a, where quantum phase fluctuations dominate as $T\rightarrow 0$ similar to underdoped La-214 cuprates\cite{Shi2014, Shi2020}.  In the weakly overdoped region (Fig.~\ref{Hall-all}b), the $R_\mathrm{H}=0$, $\rho_\mathrm{xx}\neq 0$ regime is considerably smaller, with $\delta T(H)/T_{\mathrm{c}}(H)<1$ at higher fields, indicating that here vortexlike phase fluctuations do not play an important role in the field-tuned destruction of superconductivity.  In the heavily overdoped region (Fig.~\ref{Hall-all}c), on the other hand, $\delta T(H)/T_{\mathrm{c}}(H)$ at the lowest $H$ is almost as large as in the underdoped region, but it is then quickly suppressed by the field.  However, it remains much larger than in the weakly overdoped region for all $H$.  In particular, $\delta T(H)/T_{\mathrm{c}}(H)\sim 2$, indicating that the SC transition in heavily overdoped Bi-2201 cannot be described within the BCS, mean-field treatment.  The same results for different $p$ are compared further in Fig.~\ref{Hall-all}e, where $H$ has been normalized by their respective values of $T_{\mathrm{c}}^0$.  We note that, although here $\delta T(H)/T_{\mathrm{c}}(H)$ for $p\approx 0.25$ appears to decrease towards zero as $T_{\mathrm{c}}(H)\rightarrow 0$ (see highest $H$ in Fig.~\ref{Hall-all}e) in agreement with general expectations for a phase transition driven by thermal fluctuations\cite{Sachdev2011} at a fixed $H$, it is not possible to confirm the deviation from $\delta T(H)/T_{\mathrm{c}}(H)\sim 2$ within experimental error (Fig.~\ref{Hall-all}c).  
\\
\vspace{-12pt}

\noindent\textbf{Discussion}\\
In charge- and spin-stripe-ordered  La$_{1.875}$Ba$_{0.125}$CuO$_4$ (ref.\cite{Li2019}), La$_{1.7}$Eu$_{0.2}$Sr$_{0.1}$CuO$_{4}$ and La$_{1.48}$Nd$_{0.4}$Sr$_{0.12}$CuO$_{4}$ (ref.\cite{Shi2021}), in which $p\approx 1/8$, the behavior with $R_\mathrm{H}=0$ and $\rho_\mathrm{xx}\neq 0$ was observed in the viscous VL regime, and thus attributed to the freezing or pinning of the vortex motion.  Most peculiarly, though, $R_\mathrm{H}=0$ with $\rho_\mathrm{xx}\neq 0$ was found to persist over a large range of $T$ and $H$ in the high-field normal state\cite{Li2019,Shi2021}, even with no evidence of any remnants of superconductivity\cite{Shi2021}.  While the precise origin of $R_\mathrm{H}=0$ in the normal state is still an open question, it may imply an approximate particle-hole symmetry that is unique to stripe-ordered cuprates.  Indeed, our measurements on underdoped Bi-2201 reveal that the $R_\mathrm{H}=0$ with $\rho_\mathrm{xx}\neq 0$ behavior is limited to the VL regime, i.e., it is a signature of the SC phase with nearly frozen vortexlike phase fluctuations.  These findings are reminiscent of $R_\mathrm{H}=0$ observed within the VL in some conventional disordered 2D superconductors\cite{Breznay2017, Kapitulnik2019} and oxide interfaces\cite{Chen2021}, as well as in YBa$_2$Cu$_3$O$_y$ thin films\cite{Yang2019}.  In those systems, however, $R_\mathrm{H}=0$ is found in the ``anomalous metal regime''\cite{Breznay2017, Kapitulnik2019} with $\rho_\mathrm{xx}(T\rightarrow 0)\neq 0$, in contrast to $\rho_\mathrm{xx}(T\rightarrow 0)= 0$ in Bi-2201.
It is interesting, though, that the vanishing of $R_\mathrm{H}$ in such anomalous metals has also been attributed\cite{Delacretaz2018} to the strong pinning of the vortex motion.  

Our results strongly suggest that, in Bi-2201, strong vortex pinning that gives rise to $R_\mathrm{H}=0$ with $\rho_\mathrm{xx}\neq 0$ is due to disorder.  In general, disorder is expected to lead to emergent granular superconductivity\cite{Li2021-Kiv, Bouadim2011}, i.e., the inhomogeneity of the pairing amplitude on the scale of the coherence length, forming SC puddles weakly coupled by the Josephson effect.  Although inhomogeneous, the pairing amplitude is finite throughout the system, and the SC transition occurs because of the loss of phase coherence.  The presence of intrinsic electronic inhomogeneity, such as charge order, which in Bi-2201 extends to the heavily overdoped region\cite{Peng2018, Li2021}, should enhance the tendency towards granularity and the importance of SC phase fluctuations further\cite{Kapitulnik2019}.  Scanning tunneling microscopy (STM) studies have indeed revealed evidence\cite{Pasupathy2008} for spatially inhomogeneous pairing gaps above $T_{\mathrm{c}}^{0}$ in Bi$_2$Sr$_2$CaCu$_2$O$_{8+\delta}$ (Bi-2212) near optimal doping.  At low doping, near the insulator-to-superconductor transition, both transport measurements\cite{Shi2013} on La$_{2-x}$Sr$_x$CuO$_4$ and STM studies\cite{Ye2023} on Bi-2201, have established that the onset of superconductivity, that is, of global phase coherence, is influenced by the competing charge order, and not merely by disorder.  It is thus reasonable to speculate that, in heavily overdoped Bi-2201, both disorder and charge order may affect the value of $T_{\mathrm{c}}^{0}$.  Interestingly, it has been argued that the STM results\cite{Tromp2023}, showing SC puddles above $T_{\mathrm{c}}^{0}$ in heavily overdoped Bi-2201 and at even higher $p$, beyond the superconductor-to-metal transition, cannot be described entirely within the dirty $d$-wave scenario.  This includes the extent in $T$ of pairing above  $T_{\mathrm{c}}^{0}$.  Our finding of vortexlike phase fluctuations up to $\delta T(H\rightarrow 0)/T_{\mathrm{c}}(H\rightarrow 0)\sim 7$ in heavily overdoped Bi-2201 (Fig.~\ref{Hall-all}c) is consistent with the zero-field STM evidence from that study\cite{Tromp2023} for SC puddles and pairing up to $T\sim 7~T_\mathrm{c}^0$ in Bi-2201 single crystals with the same $p \approx 0.25$.  The presence of Josephson-coupled SC grains has also been inferred\cite{Li2022} in heavily overdoped La$_{2-x}$Sr$_x$CuO$_4$.  Similar to the predictions\cite{Li2021-Kiv} for $H=0$, the dirty $d$-wave (BCS) theory shows\cite{Johnsen2023} that the interplay of disorder and orbital depairing by $H$ in hole-overdoped cuprates may lead to the emergence of an intermediate, granular regime with finite SC pairing and zero superfluid stiffness.  Our observation of pronounced phase fluctuations beyond the mean-field picture [$\delta T(H)/T_{\mathrm{c}}(H)\gg 1$] in heavily overdoped Bi-2201 is consistent with that scenario.

In summary, our comparative study of $T$--$H$ phase diagrams and low-$T$ magnetotransport properties in different doping regions of Bi-2201 using several complementary techniques has (a) demonstrated the universality of the vortex phase diagram in underdoped cuprates regardless of the presence of charge or spin orders, (b) indicated the persistence of pairing without phase coherence in an unusually wide range of $T$ above $T_{\mathrm{c}}(H)$ in the heavily overdoped region, and (c) revealed that the importance of phase fluctuations in the SC transition depends on doping in a nonmonotonic fashion.  Other experiments, such as STM studies in the presence of a magnetic field, are needed to clarify the origin of the observed enhancement of vortexlike phase fluctuations in the heavily overdoped region, as well as their possible relation to the strange-metal behavior\cite{Ayres2024}.  The latter has been indeed suggested\cite{Tranquada2024} to arise from the scattering of quasiparticles from isolated SC puddles of strongly correlated electrons.  Our results strongly suggest that it is thermal phase fluctuations that are predominantly responsible for the loss of phase coherence near the  superconductor-to-metal transition in sufficiently disordered, heavily overdoped cuprates.  Quantum fluctuations, on the other hand, play an important role in the field-tuned destruction of superconductivity in the underdoped region as $T\rightarrow 0$, but in heavily overdoped samples they are likely to become important only in the low-$T$, high-$H$ regime (cf. Fig.~\ref{fig:film}d, Fig.~\ref{fig:under}d, refs.\cite{Shi2014, Shi2020}) that is not accessible experimentally.  The quantum superconductor-to-metal transition is not well understood in general\cite{Kapitulnik2019}, and thus further insight into the role of quantum fluctuations might actually come from similar experiments on unconventional superconductors that are simpler than cuprates, such as quasi-2D organic charge transfer salts\cite{Pustogow2021}.  

\begin{methods}
\noindent\textbf{Samples}  

\noindent Several Bi-2201 samples, both single crystals and thin films, covering underdoped, overdoped, and heavily overdoped hole concentration regions were studied.\\
\vspace*{-12pt}

\noindent Single crystals of underdoped Bi-2201, Bi$_{2}$Sr$_{1.16}$La$_{0.84}$CuO$_{6+\delta}$ (BSLCO) with $p \approx 0.10$, in which La substitution for Sr is used to control hole doping concentration, were grown using the floating-zone technique\cite{Ono2003}.  The maximum $T^{0}_{\mathrm{c}}$ for the Bi$_2$Sr$_{2-x}$La$_x$CuO$_4$ family\cite{Ono2003} is $T^{0}_{\mathrm{c,max}}\sim 38$~K for $p\sim 0.16$.  Overdoped Bi$_{2.1}$Sr$_{1.9}$CuO$_{6+\delta}$ ($p\approx 0.18$) single crystals were also grown by the floating-zone technique, with $T^{0}_{\mathrm{c,max}}\sim 10$~K for Bi$_{2+x}$Sr$_{2-x}$CuO$_4$.  The hole doping concentration for both underdoped and overdoped single crystals was determined from the Hall coefficient\cite{Ando2000,Ando2001}.  The samples were shaped as rectangular bars suitable for direct measurements of the longitudinal and transverse (Hall) resistivity, $\rho_{\mathrm{xx}}$ and $\rho_{\mathrm{yx}}$, respectively.  Detailed measurements of $\rho_{\mathrm{xx}}$ and $\rho_{\mathrm{yx}}$ were performed on a $p\approx 0.10$ sample with dimensions $1.48\times 0.87\times 0.05$~mm$^3$ ($a\times b\times c$, i.e. length~$\times$~width~$\times$~thickness) and on a $p\approx 0.18$ sample with dimensions $1.27\times 0.55\times 0.03$~mm$^3$ ($a\times b\times c$). Both single crystal samples were measured in various systems, magnets and cryostats, over the period of $\sim 3$ years, and they were found to be 
stable and consistent in their behavior. The data are shown for the voltage contacts separated by 0.62~mm for Bi$_{2}$Sr$_{1.16}$La$_{0.84}$CuO$_{6+\delta}$ and 0.53~mm for Bi$_{2.1}$Sr$_{1.9}$CuO$_{6+\delta}$.\\
\vspace*{-12pt}

\noindent 
The current contacts were painted to cover two opposing side faces of the platelet-shaped crystals to ensure a uniform current flow and the voltage contacts were painted on the two remaining side faces of the crystals. The contact pads were hand-drawn with gold paint, followed by a heat treatment at 400~$^{\circ}$C for 30~min (i.e. for a week in case of a $p\approx 0.18$ crystal) in the air, which makes the gold particles adhere well to the sample surface. After this heat treatment to cure the gold contact pads, the samples were annealed in flowing air at 650~$^{\circ}$C for 48~h to control the oxygen concentration, and they were quenched to room temperature at the end of the annealing. Finally, gold wires were attached to the contact pads using EPO-TEK\textsuperscript{\textregistered} silver epoxy, which was cured at a relatively low temperature, 130~$^{\circ}$C. The resulting contact resistances were 
less than 1~$\Omega$ at room temperature\cite{Ono2003}.\\
\vspace*{-12pt}

\noindent Bi$_{2}$Sr$_{2}$CuO$_{6+\delta}$ thin films in the heavily overdoped region were grown by RF magnetron sputtering on $c$-axis oriented SrTiO3 single crystal substrates\cite{Li1993}. Oxygen annealing treatments were used to control hole doping concentration, which was determined using the phenomenological law proposed by Presland \textit{et al}.\cite{Presland1991}: $T^{0}_{\mathrm{c}}/T^{0}_{\mathrm{c,max}}=1-82.6(p-0.16)^2$, with $T^{0}_{\mathrm{c,max}}=(11.5\pm 0.5)$~K.  The films were patterned mechanically into Hall bars, parallel to the $a$ crystallographic axis, with six gold contacts for $\rho_{\mathrm{xx}}$ and $\rho_{\mathrm{yx}}$ measurements.  Detailed measurements were performed on two heavily overdoped thin films: $p\approx 0.25$ with dimensions $2\times 0.365$~mm$^2$ ($a\times b$) and thickness of 310~nm, and $p\approx 0.27$ with $2\times 0.332$~mm$^2$ ($a\times b$) and thickness of 78~nm.  The data are shown for the voltage contacts separated by 0.62~mm for the $p\approx 0.25$ sample and 0.61~mm for the $p\approx 0.27$ sample. 
\\
\vspace*{-12pt}

\noindent For each sample, $T_{\mathrm{c}}^{0}$ was defined as the temperature at which the linear resistivity becomes zero, i.e. falls below the experimental noise floor ($\sim 0.7$~m$\Omega$), giving $T_{\mathrm{c}}^{0}=(1.8\pm 0.2)$~K for $p\approx 0.10$, $T_{\mathrm{c}}^{0}=(4.8\pm 0.3)$~K for $p\approx 0.18$, $T_{\mathrm{c}}^{0}=(3.3\pm 0.2)$~K for $p\approx 0.25$, and $T_{\mathrm{c}}^{0}<0.4$~K for $p\approx 0.27$ ($T=0.4$~K was the lowest measurement temperature for this sample).  The pseudogap temperatures are $T_{\mathrm{PG}}\sim 250$~K for $p\approx 0.10$ (ref.\cite{Kawasaki2010}) and  $T_{\mathrm{PG}}\sim 100$~K for $p\approx 0.18$ (ref.\cite{Berben2022} and refs. therein).  The pseudogap closes\cite{Berben2022} at $p\sim 0.21$.\\
\vspace*{-12pt}

\noindent\textbf{Measurements}   

\noindent The standard four-probe ac method ($\sim 13$~Hz) was used for measurements of the longitudinal and transverse resistances, $R_{\mathrm{xx}}$ and $R_{\mathrm{yx}}$, respectively, with the magnetic field parallel and antiparallel to the $c$ axis.  The Hall resistance was determined from the transverse voltage by extracting the component antisymmetric in the magnetic field.  The Hall coefficient $R_{\mathrm{H}} =R_{\mathrm{yx}}\, d/H = \rho_{\mathrm{yx}}/H$, where $d$ is the sample thickness.  The resistance per square per CuO$_2$ layer $R_{\square/\mathrm{layer}}=\rho_{\mathrm{xx}}/l$, where $l=12.3$~\AA\, is the thickness of each layer.  (The $\sim 2$\% error introduced into the calculation of $R_{\square/\mathrm{layer}}$ for Bi$_{2}$Sr$_{1.16}$La$_{0.84}$CuO$_{6+\delta}$ by taking $l=12.3$~\AA\, instead of the more precise\cite{Li2005} $l\approx 12.0$~\AA\, is negligible compared to the $\sim 30$\% geometrical uncertainty in $\rho_{\mathrm{xx}}$, typical of cuprate samples\cite{Berben2022}, and it does not affect any of our conclusions.)  Depending on the temperature and magnetic field, the excitation current (density) of $1~\mu$A to $100~\mu$A ($2.3\times 10^{-3}$~A/cm$^2$ to 0.230~A/cm$^2$ for $p\approx 0.10$ sample and $6\times 10^{-3}$~A/cm$^2$ to 0.606~A/cm$^2$ for $p\approx 0.18$ sample) was used for the single crystals, and 100~nA to 300~nA (0.088~A/cm$^2$ to 0.265~A/cm$^2$) for thin films.  The excitation currents were adjusted to ensure Ohmic behavior, i.e. to avoid Joule heating.  $dV/dI$ measurements were performed by applying a dc current bias $I_{\mathrm{dc}}$ and a small ac current excitation $I_{\mathrm{ac}}$ ($\sim13$~Hz) through the sample while measuring the ac voltage across the sample. For each value of $I_{\mathrm{dc}}$, the ac voltage was monitored for 300~s, and the average value was recorded\cite{Shi2020,ZShi-PDW}.  A 1~k$\Omega$ resistor in series with a $\pi$ filter [5~dB (60~dB) noise reduction at 10~MHz (1~GHz)] was placed in each wire at the room temperature end of the cryostat to reduce the noise and heating by radiation in all measurements.
 \\
 \vspace*{-12pt}

\noindent The experiments were conducted in several different magnets at the National High Magnetic Field Laboratory: a dilution refrigerator (0.016 K $\leqslant$ T $\leqslant$ 1.1 K) in superconducting magnets with $H$ up to 18~T and 28~T, and a $^{3}$He system (0.3 K $\leqslant$ T $\leqslant$ 60 K) in a superconducting magnet with $H$ up to 18 T, using 0.1 -- 0.2~T/min sweep rates;  a $^{3}$He system (0.3 K $\leqslant$ T $\leqslant$ 15 K) in a 35~T resistive magnet, using 1 -- 2~T/min sweep rate; a $^{3}$He system (0.3 K $\leqslant$ T $\leqslant$ 30 K) in a 41~T resistive magnet, using 1~T/min sweep rate; and a $^{3}$He system (0.3 K $\leqslant$ T $\leqslant$ 12 K) in a 45~T hybrid magnet, using 1 -- 2~T/min sweep rates. The fields were swept at constant temperatures, and the sweep rates were low enough to avoid eddy current heating of the samples.  The results obtained in different magnets and cryostats agree well.
\\

\end{methods}

\noindent{\Large{\textbf{Data availability}}}  

\noindent The data that support the findings of this study are available within the paper and the Supplementary Information. Additional data related to this paper may be requested from the authors.

\bibliography{scibib}

\bibliographystyle{Science}

\vspace*{12pt}

\noindent{\Large{\textbf{Acknowledgements}}}

\noindent We acknowledge technical support from P. G. Baity and L. J.  Stanley. This work was supported by NSF Grants Nos. DMR-1707785 and DMR-2104193, and the National High Magnetic Field Laboratory through the NSF Cooperative Agreement Nos. DMR-1644779 and DMR-2128556, and the State of Florida. This research was also supported in part by the National Science Foundation under Grants No. NSF PHY-1748958 and PHY-2309135.    
S. O. acknowledges support from the JSPS KAKENHI grant (20H05304).\\
\vspace*{-6pt}

\noindent{\Large{\textbf{Author contributions}}}

\noindent Single crystals were grown and prepared by S.O.; thin films were grown and prepared by Z.Z.L. and H.R.; J.T., B.K.P., P.S. and H. R. performed the measurements;  J.T. and B.K.P. analyzed the data; J.T., B.K.P. and D.P. wrote the manuscript, with input from all authors; D.P. planned and supervised the investigation.\\
\vspace*{-6pt}

\noindent{\Large{\textbf{Competing financial interests}}}

\noindent The authors declare no competing interests.\\
\vspace*{-6pt}

\noindent{\Large{\textbf{Additional information}}}

\noindent\textbf{Supplementary information} accompanies this paper.  

\noindent\textbf{Correspondence} and requests for materials should be addressed to D.P.~(email: dragana@magnet.fsu.edu).

\clearpage

\noindent{\Large{\textbf{Supplementary Information}}}
\vspace*{0.5in}

\noindent Supplementary Figures 1-10

\clearpage

\renewcommand{\figurename}{{\bf{Supplementary Fig.}}}

\makeatletter
\makeatletter \renewcommand{\fnum@figure}{{\bf{\figurename~\thefigure}}}
\makeatother

\setcounter{figure}{0}

\baselineskip=24pt

\noindent{\large{\textbf{Supplementary Figures}}}\\
\vspace{24pt}

\begin{figure}[h]
\includegraphics[width=\textwidth]{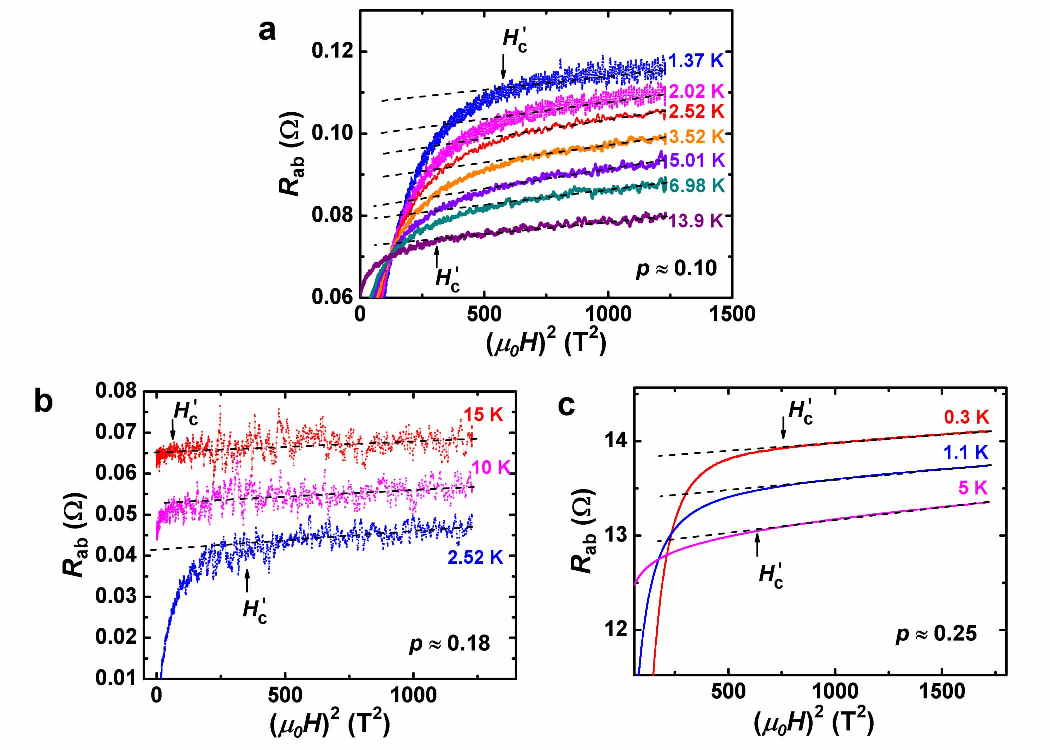}
\centering 
\caption{\textbf{In-plane resistance $\bm{R_{\mathrm{xx}}}$ vs $\bm{H^2}$ at several $\bm{T}$ for Bi-2201.}  Dashed lines are linear fits representing the contributions from normal state transport.  Deviations of the measured resistance from the fits are due to the onset of Gaussian fluctuations of the superconducting amplitude and phase, with the onset field indicated by $H_\mathrm{c}'$ for \textbf{a} underdoped ($p\approx 0.10$), \textbf{b} weakly overdoped ($p\approx 0.18$), and \textbf{c} heavily overdoped ($p\approx 0.25$) samples.  In \textbf{c}, the curves have been offset for clarity, such that the 1.1~K curve is offset by $0.4~\Omega$ and 0.3~K curve is offset by $0.8~\Omega$.
}
\label{fig:MR-H2} 
\end{figure}
%

\clearpage

\begin{figure}
\centerline{\includegraphics[width=1.08\textwidth]{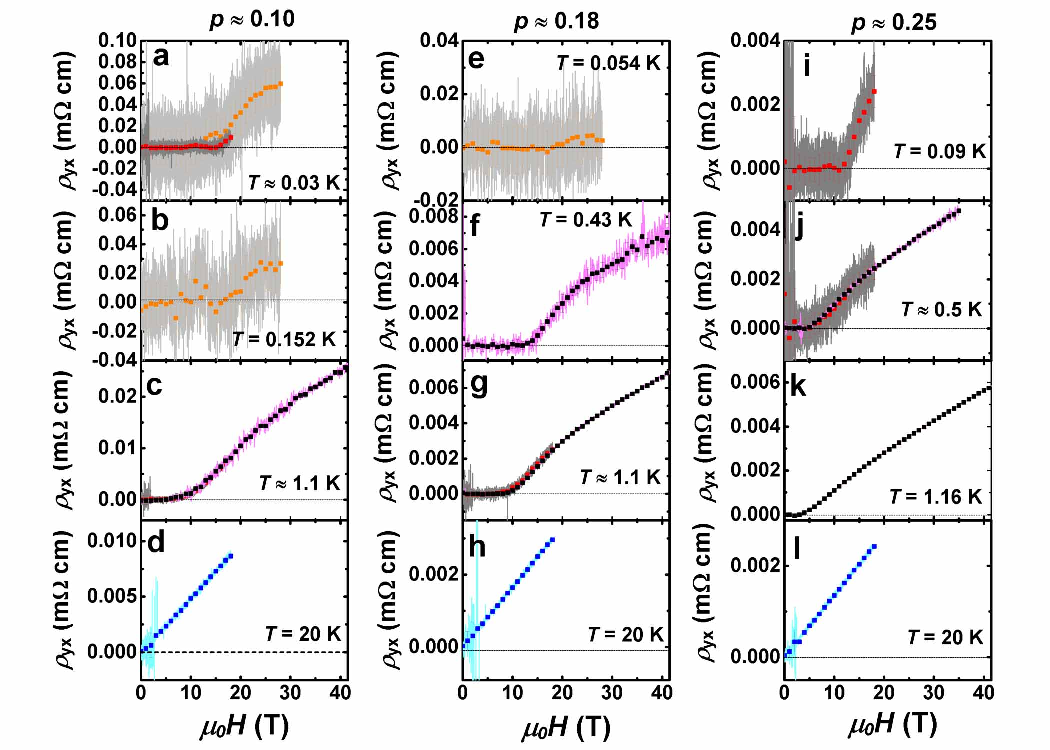}}
\caption{\textbf{Hall resistivity $\bm{\rho_{\mathrm{yx}}}$ vs $\bm{H}$ for Bi-2201.}  The data are shown for \textbf{a} - \textbf{d} $p\approx 0.10$ underdoped,   \textbf{e} - \textbf{h} $p\approx 0.18$ weakly overdoped, and \textbf{i} -\textbf{ l} $p\approx 0.25$ heavily overdoped samples.Dark gray and 
cyan traces correspond to data measured using different systems up to 18~T, while light gray and magenta traces correspond to data measured using different systems up to 28~T and 41~T, respectively. The same ${\rho_{\mathrm{yx}}}$ data, averaged over 1~T bins, are shown by red symbols (for dark gray trace), blue symbols (for cyan trace), orange symbols (for light gray trace), and black symbols (for magenta trace). Error bars correspond to $\pm$1~SD (standard deviation) of the data points within each bin. At low $T$, the signals appear relatively noisy because extremely small excitation currents $I$ are used to avoid heating and to ensure that the measurements are taken in the $I\rightarrow 0$ limit (see Methods).  At higher $T$, the signal-to-noise ratio increases.
}
\label{fig:Hall}
\end{figure}

%

\clearpage

\begin{figure}
\centerline{\includegraphics[width=1.08\textwidth]{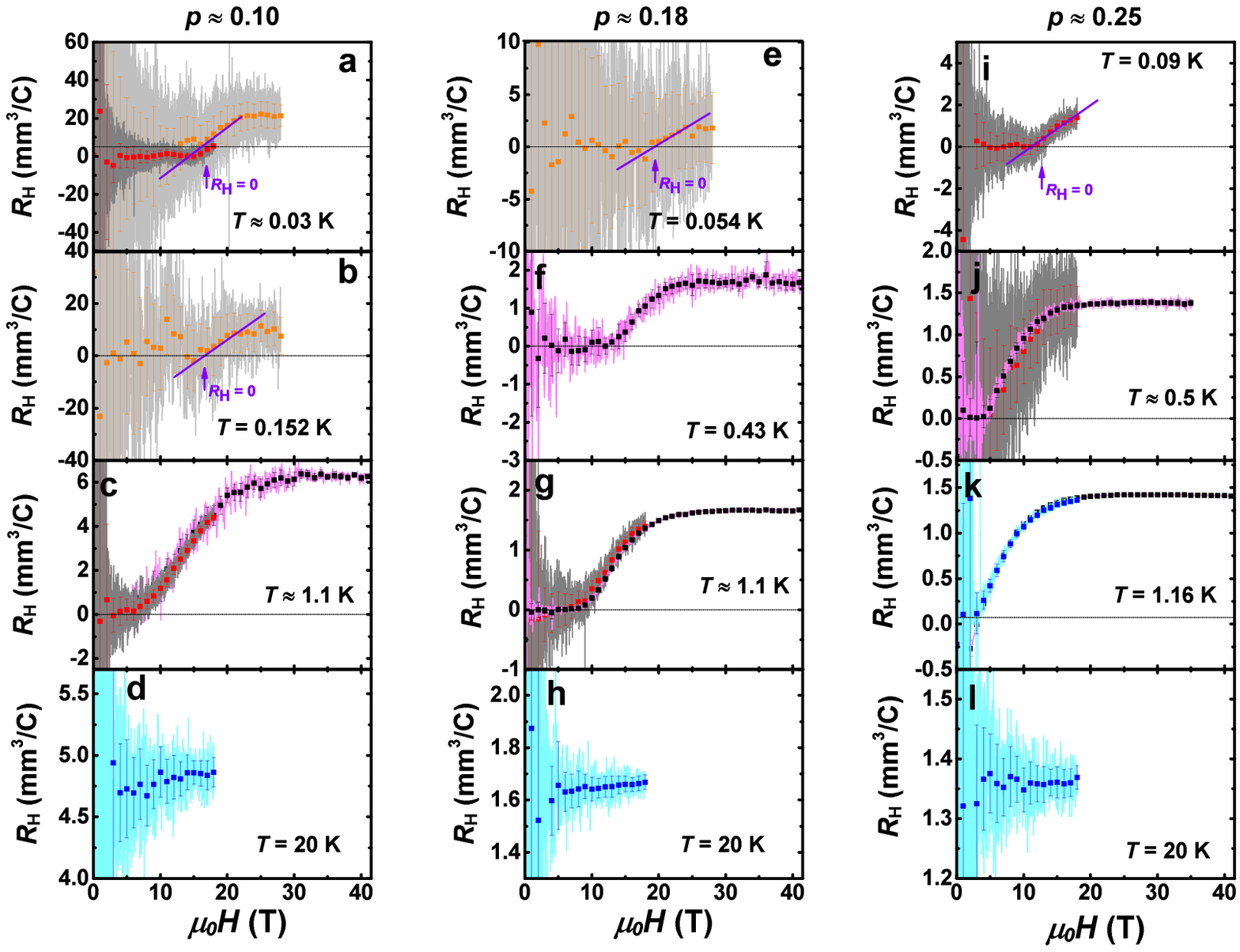}}
\caption{\textbf{Field dependence of the Hall coefficient $\bm{R_{\mathrm{H}}}$ for Bi-2201.} The data are shown for \textbf{a} - \textbf{d} $p\approx 0.10$ underdoped, \textbf{e} - \textbf{h} $p\approx 0.18$ weakly overdoped, and \textbf{i }-\textbf{l} $p\approx 0.25$ heavily overdoped samples.  Dark gray and cyan traces correspond to data measured using different systems up to 18~T, while light gray and magenta traces correspond to data measured using different systems up to 28~T and 41~T, respectively. The same $R_{\mathrm{H}}$ data, averaged over 1~T bins, are shown by red symbols (for dark gray trace), blue symbols (for cyan trace), orange symbols (for light gray trace), and black symbols (for magenta trace). Error bars correspond to $\pm$1~SD (standard deviation) of the data points within each bin. Purple lines in \textbf{a}, \textbf{b}, \textbf{e}, and \textbf{i} are linear fits to the bin-averaged data to illustrate the determination of the onset of nonzero $R_\mathrm{H}$ as the intersection with the $R_\mathrm{H}=0$ line.   
}
\label{fig:RH}
\end{figure}

\clearpage

\begin{figure}
\centerline{\includegraphics[width=0.71\textwidth]{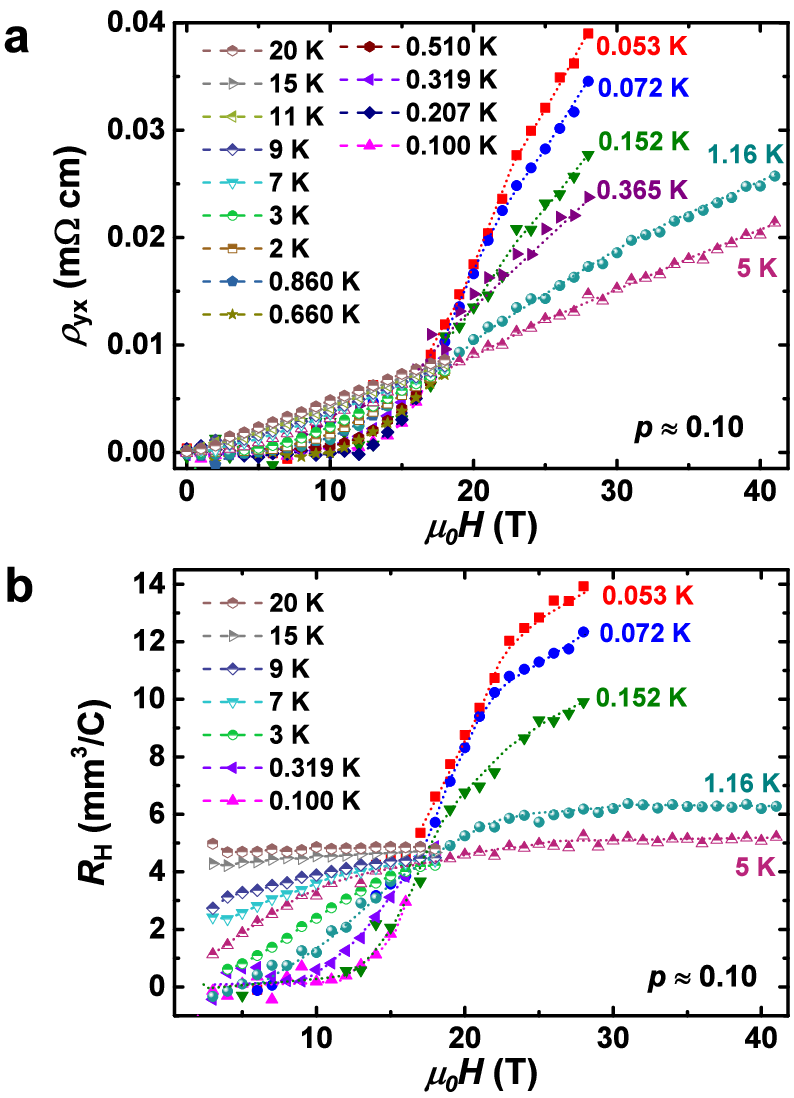}}
\caption{\textbf{Field dependence of the Hall effect in underdoped Bi-2201 (\textbf{$\bm{p\approx 0.10}$)} at various temperatures.}  \textbf{a} Hall resistivity $\rho_{\mathrm{yx}}(H)$ and \textbf{b} Hall coefficient $R_{\mathrm{H}}(H)$.  
The data points represent values obtained after averaging over 1~T bins (see Supplementary Figs.~\ref{fig:Hall} and \ref{fig:RH}).  The dotted lines guide the eye.  The error bars,  corresponding to $\pm$1~SD (standard deviation) of the data within each bin, are omitted for clarity, but they are shown in Supplementary Figs.~\ref{fig:Hall} and \ref{fig:RH}.  In \textbf{b}, $R_{\mathrm{H}}$ increases with decreasing $T$ at the highest $H$, i.e. in the normal state.  Assuming that, in the high-$H$ limit as $T\rightarrow 0$, $R_{\mathrm{H}}=1/ne$ holds ($n$ -- charge carrier density), this suggests carrier localization, consistent with weak insulating behavior observed in $\rho_{\mathrm{xx}}(T)$ in the normal state (Fig.~1a and Fig.~1c, as well as early studies$^{29}$ of underdoped Bi-2201).  
}
\label{fig:Hall-under}
\end{figure}
%

\clearpage

\begin{figure}[!b]
\centerline{\includegraphics[width=\textwidth]{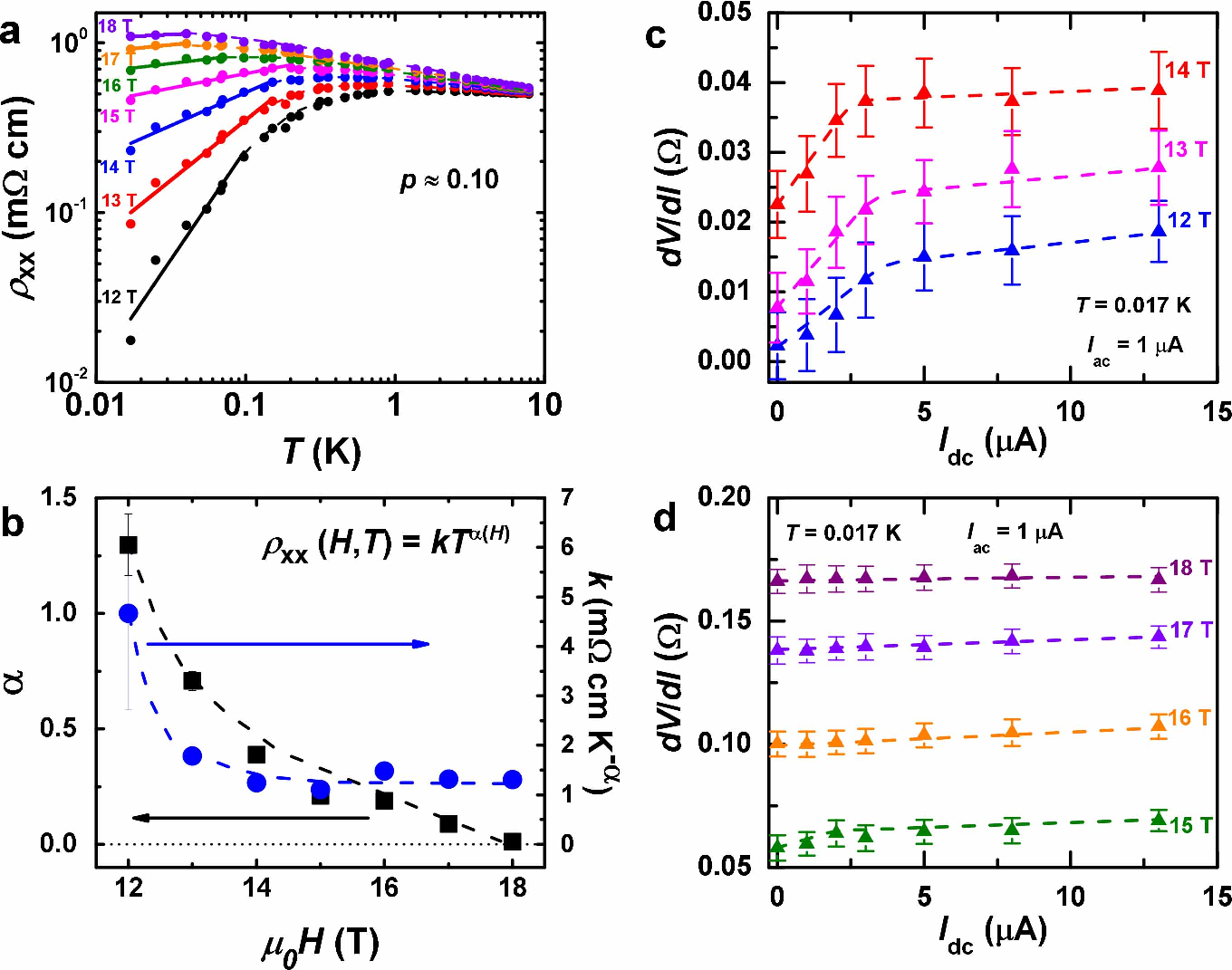}} 
\caption{\textbf{Linear and nonlinear in-plane transport at intermediate fields in underdoped Bi-2201 (\textbf{$\bm{p\approx 0.10}$)}.}  \textbf{a} Linear ($I_{\mathrm{dc}}\rightarrow 0$) resistivity $\rho_{\mathrm{xx}}(T)$ at intermediate fields.  Short-dashed lines guide the eye; solid lines represent power-law fits $\rho_{\mathrm{xx}} (H,T) = k(H)T^{\alpha(H)}$ for $T<T_{\mathrm{peak}}(H)$.  \textbf{b} Fitting parameters $\alpha$ (black squares) and $k$ (blue dots) as a function of field $H$.  Error bars correspond to $\pm~1$ SD (standard deviation) in the fits for $\alpha$ and $k$.  Dashed lines guide the eye.
\textbf{c} Differential resistance $dV/dI$ as a function of dc current $I_{\mathrm{dc}}$ for several fields at $T = 0.017$~K.  Nonlinear behavior, observed at low $I_{\mathrm{dc}}$, is expected from the motion of vortices in the presence of disorder, i.e. it is a signature of a viscous vortex liquid.  
\textbf{d} At higher $H > H^{\ast}$ (at $T=0.017$~K, $H^{\ast}\lesssim16$~T), Ohmic behavior is recovered.  In \textbf{c} and \textbf{d}, for each value of $I_{\mathrm{dc}}$ the error bar is 1~SD obtained from averaging the ac voltage over 300~s (see Methods).  Dashed lines guide the eye.
}
\label{fig:powerlaw-under}
\end{figure}

\clearpage

%
\begin{figure}
\centerline{\includegraphics[width=0.8\textwidth]{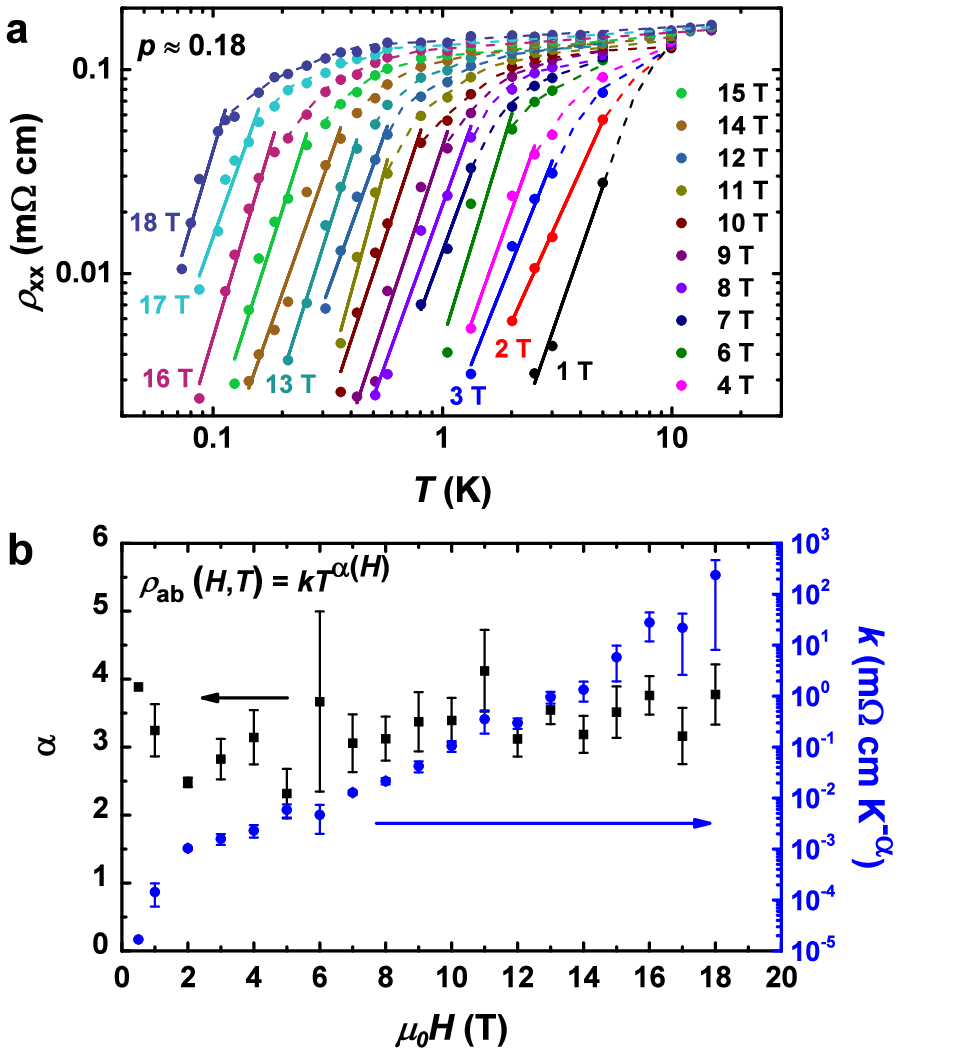}}
\caption{\textbf{Linear in-plane resistivity at several fields in weakly overdoped Bi-2201 (\textbf{$\bm{p\approx 0.18)}$}.}  \textbf{a} $\rho_{\mathrm{xx}} (T)$ for $T>T_\mathrm{c}(H)$.  The data are plotted on a log-log scale, where short-dashed lines guide the eye and the solid lines represent power-law fits ${\rho_{\mathrm{xx}}} (H,T) = k(H)T^{\alpha(H)}$. \textbf{b} Fitting parameters $\alpha$ (black squares) and $k$ (blue dots) as a function of field $H$.  Error bars correspond to $\pm~1$ SD (standard deviation) in the fits for $\alpha$ and $k$.  Although the data can be fitted with a power law, here $\alpha$ is relatively large and independent of $H$, in contrast to the behavior in the underdoped region (Supplementary Fig.~\ref{fig:powerlaw-under}a, b).
}
\label{fig:powerlaw-od}
\end{figure}
%

\clearpage

%
\begin{figure}[!tb]
\centerline{\includegraphics[width=0.7\textwidth]{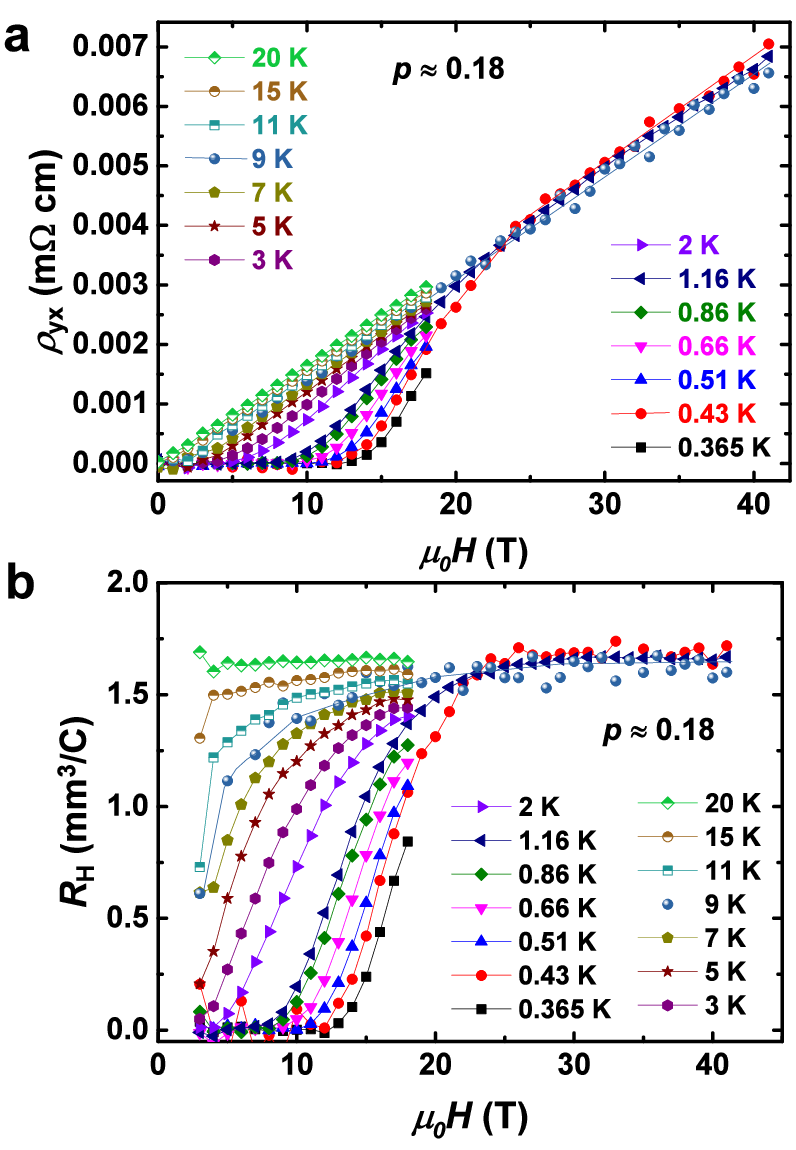}}
\caption{\textbf{Field dependence of the Hall effect in weakly overdoped Bi-2201 (\textbf{$\bm{p\approx 0.18}$)} at various temperatures.} \textbf{a} Hall resistivity $\rho_{\mathrm{yx}}(T)$ and \textbf{b} Hall coefficient $R_{\mathrm{H}}$.  The data points represent values obtained after averaging over 1~T bins (see Supplementary Figs.~\ref{fig:Hall} and \ref{fig:RH}).  The error bars, corresponding to $\pm$1~SD (standard deviation) of the data within each bin, are omitted for clarity, but they are shown in Supplementary Figs.~\ref{fig:Hall} and \ref{fig:RH}.  Thin solid lines guide the eye.  $R_{\mathrm{H}}$ becomes temperature independent for $H>H_{\mathrm{c}}'(T=0)=24$~T, where Gaussian SC fluctuations are also suppressed, signaling the onset of the normal state with the conventional metallic $\rho_\mathrm{yx}\propto H$ behavior.
}
\label{Hall-od}
\end{figure}
%

%
\begin{figure}
\centerline{\includegraphics[width=0.9\textwidth]{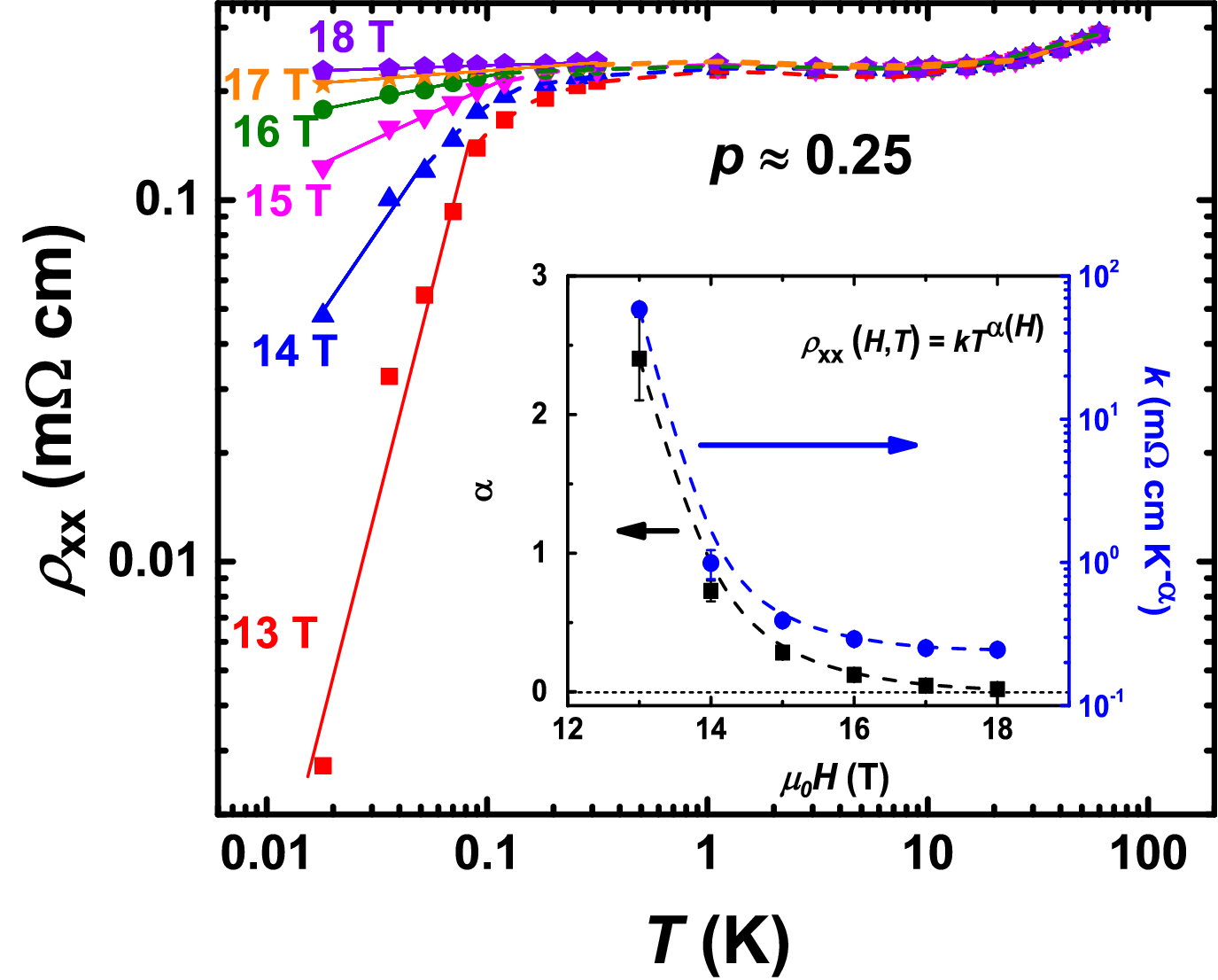}}
\caption{\textbf{Linear in-plane resistivity at intermediate fields $\bm{H}$ in heavily overdoped Bi-2201 (\textbf{$\bm{p\approx 0.25}$)}.}
$\rho_{\mathrm{xx}} (T)$ is plotted on a log-log scale where short-dashed lines guide the eye and the solid lines represent power-law fits ${\rho_{\mathrm{xx}}} (H,T) = kT^{\alpha(H)}$ for $T<T_{\mathrm{peak}}(H)$.  Inset: Power-law fitting parameters $\alpha$ (black squares) and $k$ (blue dots) as a function of field $H$.  Error bars correspond to $\pm~1$ SD (standard deviation) in the fits for $\alpha$ and $k$.  Dashed lines guide the eye.
}
\label{film-power}
\end{figure}

\clearpage

\begin{figure}
 \centering
    \includegraphics[width=0.9\textwidth]{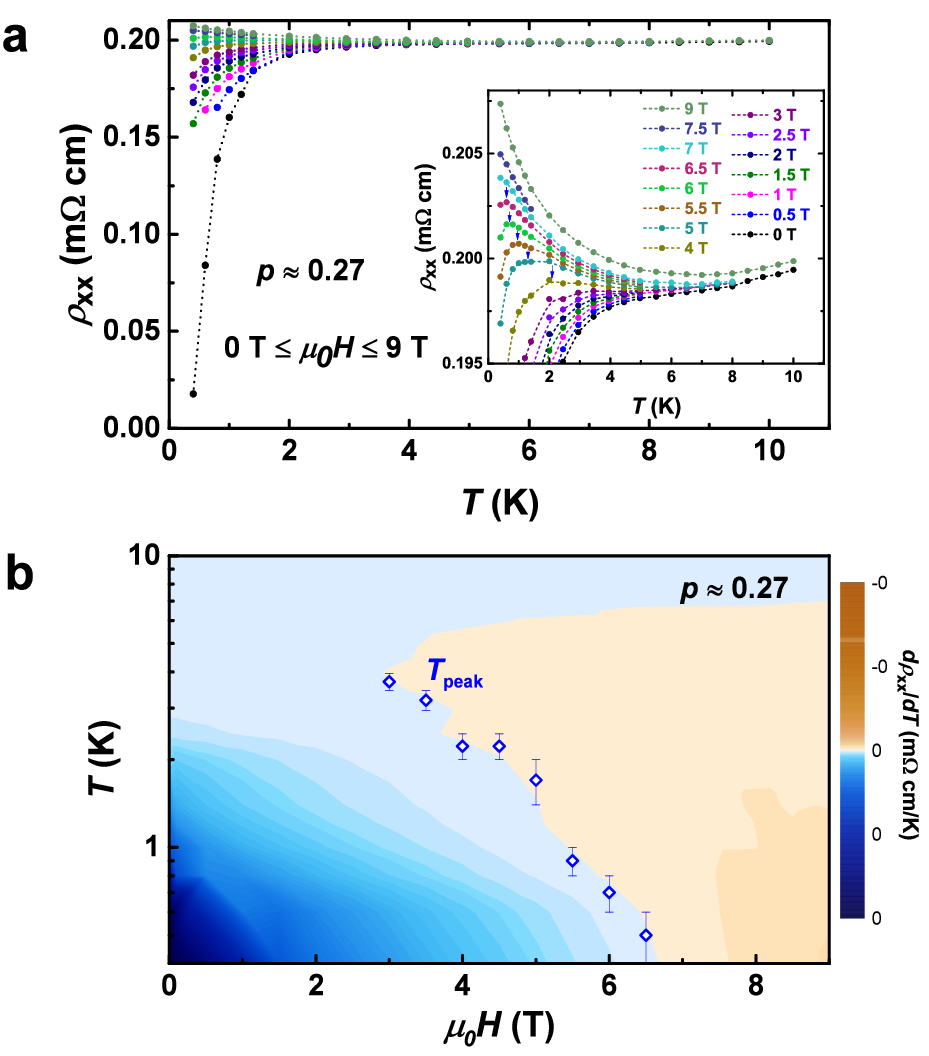}
\caption{\textbf{Linear in-plane resistivity in heavily overdoped Bi-2201 (\textbf{$\bm{p\approx 0.27}$)}.}  \textbf{a} $\rho_{\mathrm{xx}}(T)$ for several 
$H\leq 9$~T; $H\parallel c$.  Inset: Some of the same data zoomed-in to emphasize $T_{\mathrm{peak}}$ marked by arrows.  \textbf{b} $T$--$H$ phase diagram; color map: $d\rho_{\mathrm{xx}}/dT$. $T_{\mathrm{peak}}$ (open blue diamonds) mark the position of the peak in $\rho_{\mathrm{xx}}(T)$.  The lowest measurement temperature for this sample was $T=0.4$~K, so $T_{\mathrm{c}}^{0}<0.4$~K.  
}
\label{B-B1}
\end{figure}

\clearpage

\begin{figure}
\centerline{\includegraphics[width=0.9\textwidth]{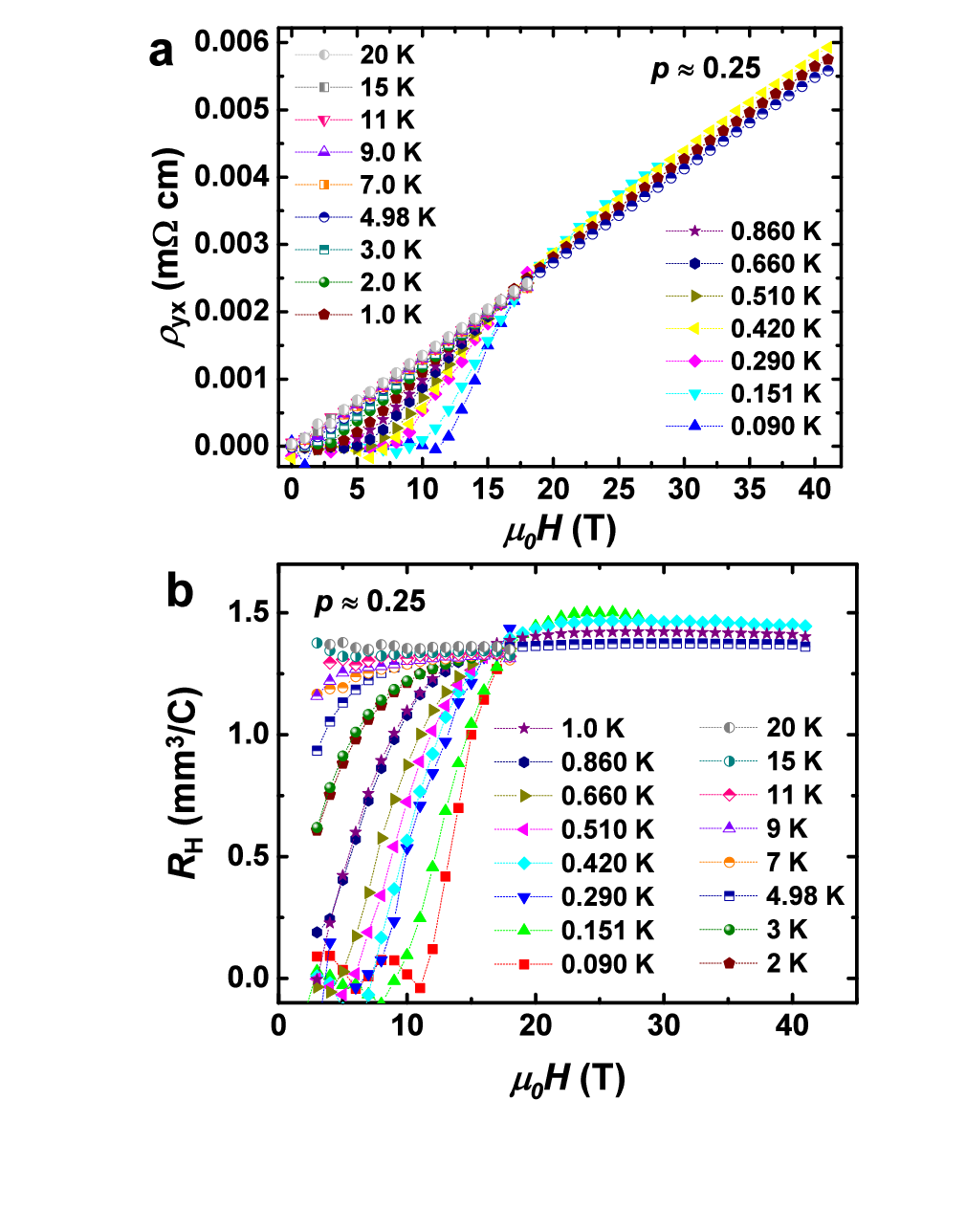}}
\vspace*{-48pt}
\caption{\textbf{Field dependence of the Hall effect in heavily overdoped Bi-2201 (\textbf{$\bm{p\approx 0.25}$)} at various temperatures.} 
\textbf{a} Hall resistivity $\rho_{\mathrm{yx}}(T)$ and \textbf{b} Hall coefficient $R_{\mathrm{H}}$.  The data points represent values obtained after averaging over 1~T bins (see Supplementary Figs.~\ref{fig:Hall} and \ref{fig:RH}).  The error bars, corresponding to $\pm$1~SD (standard deviation) of the data within each bin, are omitted for clarity, but they are shown in Supplementary Figs.~\ref{fig:Hall} and \ref{fig:RH}.  Thin solid lines guide the eye.  
}
\label{B1}
\end{figure}
%

\end{document}